# New Compensating and Equivalent Variation Closed-form Solutions for Non-Separable Public Goods

Daniel H. Karney, Department of Economics, Ohio University

Khyati Malik, National Bureau of Economic Research (NBER)

January 2026

**Abstract.** This study finds exact closed-form solutions for compensating variation (CV) and equivalent variation (EV) for both marginal and non-marginal changes in public goods given homothetic, but non-separable, utility where a single sufficient statistic summarizes consumer preferences. The closed-form CV and EV expressions identify three economic mechanisms that determine magnitudes. One of these mechanisms, the relative preference effect, helps explain the disparity between willingness to pay (WTP) and willingness to accept (WTA) for public goods. We also show how our closed-form solutions can be employed to calculate WTP and WTA across income groups using estimates from existing empirical studies.

KEYWORDS: Public goods; Compensating variation; Equivalent variation; Willingness-to-pay; Willingness-to-accept; Non-separable utility

---

Contacts: Daniel H. Karney: karney@ohio.edu; Khyati Malik: malikk@nber.org

We are grateful for comments from Marcel Arbex, Spencer Banzhaf, Peter Caradonna, Nathan Chan, Sathya Gopalakrishnan, Anna Klis, Bethany Lemont, Vahe Lskavyan, James Price, and from participants at 2024 AERE@MEA and 2024 AERE Summer Conference. This manuscript was previously circulated under the title "Public Good Provision and Compensating Variation". All mistakes are our own.

# 1 Introduction

Public goods provision by governments is a prominent feature of modern economies. In 2019, U.S. local municipalities and states spent $425.7 billion on gross investments in fixed assets including streets, highways, and water and sewer systems (U.S. BEA, 2022). The U.S. federal government spends hundreds of billions of dollars on public goods like national defense (U.S. GAO, 2021) as well as provides other public goods via policies that incur costs without direct expenditures such as mandates to reduce pollution.[1] Projected costs of climate change mitigation may be one of the largest future public goods expenditure as worldwide total costs are expected to range from one to seven percent of global GDP (Fujimori *et al.*, 2023). How much are consumers willing to pay for these public goods? Compensating variation (CV) and equivalent variation (EV) measure the value of public good provision (Hanemann, 1991; Weber, 1992).

This study contributes to the literature by finding closed-form solutions for CV and EV for changes in public good provision. We also find a single sufficient statistic for the preference parameters and show how to estimate this sufficient statistic in empirical applications. Previous studies to uncover willingness to pay (WTP) and willingness to accept (WTA) of public goods often impose separability to generate tractable results such as strongly or weakly separable utility (Freeman *et al.,* 2014). Here, we study a utility model with non-separable private and public goods with homothetic preferences where public good provision may affect private consumption choices on the margin. For instance, the provision of clean municipal water allows consumers to substitute away from private adaptation goods such as water filtration devices. Meanwhile, the building of public highways complements private vehicle purchases. Hence, the private and public goods may be complements or substitutes for each other in our model.

[1] Present government expenditures towards public goods may not change current public good provision. For example, an under-construction municipal water system may entail significant upfront costs to the government but will not contribute to an individual's current-period direct utility in our model until it starts supplying clean water.

Other assumptions in the literature to enable tractability in the context of valuing public goods include weak complementarity (Mäler, 1974), the Willig condition (Willig, 1978; Smith and Banzhaf, 2004; Palmquist, 2005), and weak substitution (Feenberg and Mills, 1980). In general, this study adds to the extensive theoretical (e.g., Larson, 1992; Neill, 1998) and empirical (e.g., Ito and Zhang, 2020; Banzhaf, 2021) literature on willingness to pay for public goods and local amenities. Our results also relate to the theory of non-traded goods (Neary and Roberts, 1980; Bockstael and McConnell, 1993). Finally, this analysis extends the work on homogeneous utility functions but without public goods (Espinosa and Prada, 2012).

We derive conditions for recovering closed-form expressions of compensating variation and equivalent variation for utility functions with homothetic preferences for a change in public good provision. The family of homothetic utility functions that we consider allow private and public goods to be independently homogeneous in the underlying function without restrictions on the monotonic transformation. We define "independently homogeneous" to mean that a function is homogeneous in the private goods and separately in the public goods of (potentially) different degrees.[2]

Homothetic utility functions are widely used in economic theory and applied applications, and common forms include the constant elasticity of substitution (CES) and constant relative risk aversion (CRRA) utility functions. As a result, homothetic utility functions are employed in virtually all fields of economics including, but not limited to, consumer theory (Caselli and Ventura, 2000), international trade (Kucheryavyy, 2012), and political economy (Collie, 2022). In the context of environmental economics per our numerical application, homothetic functions have been employed in numerical work (e.g., Pindyck, 2012), applied theory exercises (e.g., Fullerton and Chi, 2019), as well as computational general equilibrium studies (e.g., Goulder and Hafstead, 2017).

[2] In contrast, a "jointly homogeneous" function is homogeneous in both private and public goods simultaneously to the same degree across both arguments.

The assumption of homotheticity with underlying independent homogeneity is crucial in our approach because it leads to exact closed-form expressions for welfare measures—specifically, compensating and equivalent variation—without resorting to consumer surplus approximations. By exploiting these closed-form expressions, we can systematically link changes in public goods provision to income in a way that remains analytically tractable. This is significant not just from a theoretical standpoint; in practice, it allows researchers and policy analysts to directly compute welfare measures, sidestepping the limitations that can arise from relying on approximate measures of consumer surplus.

The single sufficient statistic in our CV and EV closed-form solutions is the ratio of the degrees of homogeneity of the public and private goods from the independently homogeneous function. This ratio is a sufficient statistic because it reduces the degree of complexity in our CV and EV expressions from two parameters to one parameter (Reid, 2015), and the statistic indicates the sign and relative size of CV and EV, all else equal (Chetty, 2009). Also, the sufficient statistic can be estimated using either individual- or aggregate-level data of consumer expenditure (or equivalently income), or private goods purchases given variation in public good provision. Our utility function may also include an auxiliary set of public goods that are not subject to homogeneity assumptions, and therefore not all public goods must have the same homothetic relationship to the private goods. Only a few studies explore willingness to pay for public goods given homothetic utility (e.g., Quigley, 1982; Chattopadhyay, 1999) and ours is the first study to derive these specific closed-form CV and EV expressions.

Our closed-form expressions allow us to cleanly identify economic mechanisms driving CV and EV. We identify three different effects and relate them to WTP (the intuition is similar for WTA). First, larger initial income leads to a larger WTP, all else equal, and this is the income effect. Second, the absolute preference effect finds a positive correlation between the preference parameter governing the impact of public good provision in utility and WTP. Third, the relative preference effect leads to a larger WTP if the ratio of the degrees of homogeneity of the public goods and the private goods is large. Conversely, the relative preference effect can result in a small WTP if the ratio is small, and this relative effect is independent of the level of the public good preference parameter. Thus, the relative preference effect demonstrates how WTP for public goods can be small, regardless of income level and absolute effect of public goods in utility.

Our relative preference effect is related to the substitution effect discussed in Hanemann (1991), but we show that the independent homogeneity assumption requires the private and public goods to be imperfect substitutes. While Hanemann (1991)'s main objective is to show how substitution could generate a large gap between WTP and WTA even in a traditional framework, much of the experimental and survey literature has emphasized behavioral explanations, such as loss aversion (Kahneman, Knetsch, and Thaler, 1990). We demonstrate that in a neoclassical model with homothetic (non-separable) preferences without any behavioral biases, a gap can arise via the relative preference effect. This effect is consistent with a substitution elasticity that drives a wedge between WTP and WTA for public goods and helps clarify that WTP–WTA differences need not rely solely on behavioral departures from the neoclassical assumptions. We prove that the relative preference effect partially drives the WTP-WTA gap for public goods in our model, and this can help explain the Tunçel and Hammitt (2014) finding of a consistently larger gap between WTP and WTA for public goods compared to private goods. Furthermore, our closed-form CV and EV expressions are asymmetric, and this matches the asymmetry between WTP and WTA found in the empirical literature (Horowitz and McConnell, 2002).

Also, while many WTP and WTA results hold for only marginal changes, our CV and EV closed-form solutions hold for both marginal and non-marginal changes in public good provision. Non-marginal changes in public good provision occur with large-scale public investments, such as the construction of the U.S. interstate highways (Duranton and Turner, 2012) or implementation of a paradigm-changing public policy, such as the Clean Air Act (Aldy *et al.,* 2022). Therefore, our study also contributes to the literature by providing an exact measure of WTP and WTA for non-marginal public good provision changes. Again, our CV and EV expressions do not rely on consumer surplus to approximate welfare as occurs with other WTP/WTA measures including those employing the Willig condition.[3] Slesnick (1998) discusses a general set of issues related to using consumer surplus as an approximation for individual welfare in a variety of settings.

As an illustration of our results, we draw on existing studies—such as Ito and Zhang's (2020) analysis of air purifier markets in China to demonstrate that observed WTP for pollution reductions can be readily translated into the sufficient statistic from our CV and EV closed-form expression. We then show how to use our closed-form expressions to easily calculate WTP (or WTA) across income groups using standard empirical estimates.

---

[3] With regards to the welfare effects of public goods and valuing public goods, Bockstael and McConnell (1993, p.1254) note, "The Willig condition describes what must be true for the incremental consumer surplus to equal the marginal value of quality measured from the expenditure function. When the Willig condition holds, the value of a change in quality measured by a change in consumer surpluses for the associate good will be bounded by the equivalent and compensating variation from a quality change."

This paper serves as a bridge between theory and empirical application by offering an analytically precise method to calculate welfare metrics—such as compensating and equivalent variation—for public goods under homothetic, non-separable preferences. In doing so, it departs from purely theoretical modeling in two keyways. First, the closed-form CV and EV expressions make it straightforward to connect the degree of homotheticity to observed WTP or WTA in applied contexts. Second, because our expressions hold for non-marginal as well as small changes in public goods, this framework is immediately usable by environmental economists, public finance analysts, and others engaged in cost–benefit evaluations that often extend beyond incremental changes. By eliminating the need for consumer-surplus approximations, our approach offers a direct, exact mapping from measurable changes in public-goods provision to underlying welfare gains or losses. This provides a practical tool for researchers and policymakers who require valuations of large-scale public investments.

The next section details the theoretical framework and accompanying assumptions. Section 3 provides our main results including CV and EV closed-form solutions via the expenditure function. Next, Section 4 calculates values of the sufficient statistic from empirical WTP estimates in the literature and then shows how to find WTP estimates for different income groups using the derived sufficient statistic. Finally, Section 5 compares our results to the common assumption of additively separable utility, and Section 6 concludes.

# 2 Model

Given (potentially) many non-separable public goods, let the representative consumer's utility maximization problem (UMP) be defined as $\max_{\boldsymbol{x}}\{h(\boldsymbol{x};\boldsymbol{z},\boldsymbol{w}): \boldsymbol{p}\cdot\boldsymbol{x} \le m\}$, where $\boldsymbol{x}\in\mathbb{R}_{+}^{N}$ is a row vector of private goods indexed $n = 1,\dots,N$ with associated prices $\boldsymbol{p}\in\mathbb{R}_{++}^{N}$ and exogenous income $m > 0$. Let $\boldsymbol{y} = (\boldsymbol{z},\boldsymbol{w})\in\mathbb{R}_{+}^{K}$ be a row vector of public goods indexed as $k = 1,\dots,K$ with $\boldsymbol{z}\in\mathbb{R}_{+}^{K_1}$ and $\boldsymbol{w}\in\mathbb{R}_{+}^{K_2}$ where $K = K_1 + K_2$. The function $h(\cdot)$ is a homothetic utility function defined as $h(\boldsymbol{x};\boldsymbol{z},\boldsymbol{w}) \equiv g\big(u(\boldsymbol{x};\boldsymbol{z},\boldsymbol{w})\big)$ where $g(\cdot)$ is a monotonic transformation and $u(\cdot)$ is the so-called "underlying" function that is either jointly or independently homogeneous in $\boldsymbol{x}$ and $\boldsymbol{z}$. Assume $u(\cdot)$ is strictly increasing, continuous, and quasiconcave.

Private goods are purchased via marginal pricing. Public goods such as clean municipal water or a highway network, provide utility to the consumer but are not purchased.[4] The level of public good provision is exogenous from the consumer's perspective so $\boldsymbol{y}$ is not included in the consumer's budget constraint. For notational purposes, $\boldsymbol{y}$ is offset by a semicolon in the relevant functions to denote exogeneity. A change in the level of public good provision may occur through quantity change and/or quality change but what matters to the consumer is the flow of utility-generating public goods. That is, the public goods vector represents the flow of value in utility and not the physical stock of public good infrastructure.

[4] The goods in vector $\boldsymbol{y} = (\boldsymbol{z}\,\boldsymbol{w})$ do not necessarily need to be non-rival and non-excludable and could be rival or excludable, but goods provided freely to consumers by governments are understood to be public goods in many cases. Public good provision may be voluntarily provided such as vaccine uptake, but our model does not allow for impure public goods whereby purchasing $\boldsymbol{x}$ affects public good provision $\boldsymbol{y}$ (Chan and Kotchen, 2014; Chan, 2024). Also, the model does not allow the representative consumer to select the level or composition of public good provision via Tiebout (1956) sorting.

The model does not require optimal public good provision because the consumer optimizes utility and private consumption regardless of the level of public goods provided. The UMP leads to Marshallian demand $\boldsymbol{x}(m, \boldsymbol{p}; \boldsymbol{z}, \boldsymbol{w})$ and the subsequent indirect utility function $v(m, \boldsymbol{p}; \boldsymbol{z}, \boldsymbol{w})$. The dual expenditure minimization problem (EMP), formally given by $\min_x\{\boldsymbol{p} \cdot \boldsymbol{x}: h(\boldsymbol{x}; \boldsymbol{z}, \boldsymbol{w}) \geq u\}$, yields Hicksian demand, $\boldsymbol{x}^{\boldsymbol{h}}(u, \boldsymbol{p}; \boldsymbol{z}, \boldsymbol{w})$, and the expenditure function, $e(u, \boldsymbol{p}; \boldsymbol{z}, \boldsymbol{w})$, where $u$ is a scalar, fixed utility level.

We define three general characteristics of the utility function. First, the utility function is strictly increasing in the public and private goods. Second, the utility function is continuous and quasi-concave. Third, the utility function is homothetic and the underlying function, $u(\cdot)$, is homogeneous. For the special case of homogeneous utility, one would employ $u(\boldsymbol{x}; \boldsymbol{z}, \boldsymbol{w})$ directly as the utility function in the consumer's optimization problem. However, our main results for CV and EV employ a homothetic utility function given by $h(\boldsymbol{x}; \boldsymbol{z}, \boldsymbol{w}) \equiv g\big(u(\boldsymbol{x}; \boldsymbol{z}, \boldsymbol{w})\big)$ so that the CV and EV results only use the ordinal properties of the underlying $u(\cdot)$ function.

For our main results, we define $u(\cdot)$ to be "independently homogeneous", or exhibits "independent homogeneity", across the vectors $\boldsymbol{x}$ and $\boldsymbol{z}$ where the function is homogeneous of degree $\eta \in \mathbb{R}$ in $\boldsymbol{x}$ and homogeneous of degree $\theta \in \mathbb{R}$ in $\boldsymbol{z}$; that is, $t^{\eta} u(\boldsymbol{x}; \boldsymbol{z}, \boldsymbol{w}) = u(t\boldsymbol{x}; \boldsymbol{z}, \boldsymbol{w})$ for all $t > 0$ and $t^{\theta} u(\boldsymbol{x}; \boldsymbol{z}, \boldsymbol{w}) = u(\boldsymbol{x}; t\boldsymbol{z}, \boldsymbol{w})$ for all $t > 0$. It follows that $t^{\eta+\theta} u(\boldsymbol{x}; \boldsymbol{z}, \boldsymbol{w}) = u(t\boldsymbol{x}; t\boldsymbol{z}, \boldsymbol{w})$ such that an independently homogeneous function is also "jointly homogeneous" of degree $\eta + \theta$. The standard homogeneity and homotheticity results then apply in the cone $(\boldsymbol{x}\ \boldsymbol{z}) \in \mathbb{R}_{+}^{N+K_1}$ as well as the cones defined by the clauses of the independent homogeneity assumption (Simon and Blume, 1994). As a result, there always exists an interior solution with positive prices for the private goods since the income expansion rays are linear from the origin.

In contrast, one could define $u(\cdot)$ to be a jointly homogeneous function of degree $\gamma$ in $\boldsymbol{x}$ and $\boldsymbol{z}$; that is, $t^{\gamma}u(\boldsymbol{x};\boldsymbol{z},\boldsymbol{w}) = u(t\boldsymbol{x};t\boldsymbol{z},\boldsymbol{w})$ for all $t > 0$ where $\gamma \in \mathbb{R}$ denotes the degree of homogeneity. In such cases, we would say the function exhibits "joint homogeneity", since it is homogeneous in the private goods and the public goods to the same degree and each argument must be scaled by the same $t > 0$.

The second public goods vector $(\boldsymbol{w})$ need not be included in any of the homogeneity conditions and thus represents an auxiliary public goods vector. We include this auxiliary vector in our model to highlight that not all public goods need to be subject to a homogeneity assumption to yield results. Also, the first vector of public goods $(\boldsymbol{z})$ covers the special cases of a single, scalar public good $(z)$ as well as an aggregator function that translates a vector of public goods into a single-valued measure (e.g., $z = f(\boldsymbol{z})$).

We highlight that our utility model allows for a flexible relationship between private and public goods without assuming strongly or weakly separable utility. Freeman *et al.* (2014) defines a strongly separable utility function where the marginal rate of substitution (MRS) of any two goods, regardless of the subset, is independent of the quantity of any other good. Weakly separable utility function means that the MRS for any pair of goods within a subset of goods is independent of the quantities in other subsets. We make no such assumptions. Our modeling assumptions also imply that there are no choke prices for any of the private goods meaning that weak complementarity does not hold (Smith and Banzhaf, 2004). Relatedly, but independently, the Willig condition is not implicit in our model either as Palmquist (2005)'s path-independence condition does not hold.

# 3 Compensating and Equivalent Variation

Using the indirect utility function, CV and EV are implicitly defined in Hanemann (1991) as $v(m - \text{CV}, \boldsymbol{p}; \boldsymbol{z}_1, \mathbf{w}) = v(m, \boldsymbol{p}; \boldsymbol{z}_0, \mathbf{w}) = u_0$ and $v(m, \boldsymbol{p}; \boldsymbol{z}_1, \mathbf{w}) = v(m + \text{EV}, \boldsymbol{p}; \boldsymbol{z}_0, \boldsymbol{w}) = u_1$, respectively, with $\boldsymbol{z}_0, \boldsymbol{z}_1 \in \boldsymbol{z}$ where $\boldsymbol{z}_1 \equiv t\boldsymbol{z}_0$ such that $\boldsymbol{z}_0$ is the initial value of $\boldsymbol{z}$, while $\boldsymbol{z}_1$ is the new, scaled value. CV and EV are both positive (negative) when $t > 1$ ($0 < t < 1$). Theorem 1 is our main result.[5,6]

**Theorem 1.** Let $g(\cdot)$ be a monotonic transformation such that utility function $h(\boldsymbol{x}; \boldsymbol{z}, \boldsymbol{w}) \equiv g\big(u(\boldsymbol{x}; \boldsymbol{z}, \boldsymbol{w})\big)$ is homothetic. If the function $u(\boldsymbol{x}; \boldsymbol{z}, \boldsymbol{w})$ is independently homogeneous, such that $t^{\eta} u(\boldsymbol{x}; \boldsymbol{z}, \boldsymbol{w}) = u(t\boldsymbol{x}; \boldsymbol{z}, \boldsymbol{w})$ with $\eta \neq 0$ for all $t > 0$ and $t^{\theta} u(\boldsymbol{x}; \boldsymbol{z}, \boldsymbol{w}) = u(\boldsymbol{x}; t\boldsymbol{z}, \boldsymbol{w})$ for all $t > 0$, then:

(a) Expenditure function $e(u, \boldsymbol{p}; \boldsymbol{z}, \boldsymbol{w})$ related to the utility function $h(\boldsymbol{x}; \boldsymbol{z}, \boldsymbol{w})$ has the property $t^{-\theta/\eta} e(u, \boldsymbol{p}; \boldsymbol{z}, \boldsymbol{w}) = e(u, \boldsymbol{p}; t\boldsymbol{z}, \boldsymbol{w})$ for all $t > 0$, that is, the expenditure function is homogeneous of degree $\phi \equiv (-\theta/\eta)$ in the $\boldsymbol{z}$ vector of public goods;

(b) Hicksian demand for the utility function $h(\boldsymbol{x}; \boldsymbol{z}, \boldsymbol{w})$ is homogenous of degree $\phi$ in public goods ($\boldsymbol{z}$); that is, $t^{\phi} x_n^h(u, \boldsymbol{p}; \boldsymbol{z}, \boldsymbol{w}) = x_n^h(u, \boldsymbol{p}; t\boldsymbol{z}, \boldsymbol{w})$ for all $t > 0$ for all $n = 1, \dots, N$, when $u(\boldsymbol{x}; \boldsymbol{z}, \boldsymbol{w})$ is strictly quasiconcave; and,

---

[5] Theorem 1 requires $\eta \neq 0$ because $\eta = 0$ means the degree of homogeneity for the expenditure function is undefined.

[6] Since any rational number can be written as a fraction, then $\phi$ as a rational number implies the existence of integer values for $\theta$ and $\eta$. The ratio $\theta/\eta$ is independent of scale since multiplying the numerator and denominator by $1/\zeta$, where $\zeta$ is a non-zero integer, yields the same $\phi$. Therefore, Theorem 1 parts (a)-(b) provide a family of utility functions necessary for a homogeneous expenditure and Hicksian demand function, respectively.

(c) Compensating variation (CV) and equivalent variation (EV) related to the utility function $h(\boldsymbol{x}; \boldsymbol{z}, \boldsymbol{w})$ are given by the expressions $\text{CV}(\boldsymbol{z_0}, t) = e(u_0, \boldsymbol{p}; \boldsymbol{z_0}, \boldsymbol{w}) - e(u_0, \boldsymbol{p}; t\boldsymbol{z_0}, \boldsymbol{w})$ and $\text{EV}(\boldsymbol{z_0}, t) = e(u_1, \boldsymbol{p}, \boldsymbol{z_0}, \boldsymbol{w}) - e(u_1, \boldsymbol{p}; t\boldsymbol{z_0}, \boldsymbol{w})$. These expressions are equivalent to $\text{CV}(m, t) = (1 - t^{\phi})m$ and $\text{EV}(m, t) = (t^{-\phi} - 1)m$, respectively, where $m$ is income.

**Proof of Theorem 1.** All main text proofs appear in Appendix A.

Theorem 1 part (a) proves that a homothetic utility function with an independently homogeneous underlying function leads to a homogeneous expenditure function in public goods $(\boldsymbol{z})$. The expenditure function is homogeneous of degree $(-\theta/\eta) \equiv \phi$ and thus equal to the negative ratio of the degrees of homogeneity for the public and private goods. Theorem 1 part (b) immediately follows from the expenditure function result and shows that Hicksian demand is homogeneous of degree $(-\theta/\eta) \equiv \phi$ in the public goods $(\boldsymbol{z}_1)$ when $u(\cdot)$ is strictly quasiconcave.

Then, Theorem 1 part (c) provides the closed-form CV and EV expressions given by $\text{CV}(m, t) = (1 - t^{-\theta/\eta})m$ and $\text{EV}(m, t) = (t^{\theta/\eta} - 1)m$. For an increase in public good provision $(t > 1)$, the closed-form CV and EV expressions demonstrate three effects that impact WTP and WTA, respectively, where $\phi \equiv (-\theta/\eta) < 0$ is a sufficient statistic summarizing preferences. First, the income effect shows that CV and EV increase with income. Second, the absolute preference effect occurs when CV and EV increase in the public good preference parameter $\theta$, holding income and $\eta$ constant. Third, the relative preference effect identifies that CV and EV increase in the ratio $\theta/\eta$ and this trade-off in utility is independent of proportionally scaling $\theta$ and $\eta$ whereby CV and EV can be small even if $\theta$ is "large" (noting $\theta/\eta$ is not generally equal to the marginal rate of substitution between $\boldsymbol{x}$ and $\boldsymbol{z}$).

For a decrease in public good provision $(0 < t < 1)$ such that CV and EV are negative (with $\mathrm{CV} = \mathrm{EV} = 0$ when $t = 1$), the resulting welfare losses imply that CV is equivalent to WTA while EV is equivalent to WTP. However, relevant to our numerical application below, when $\boldsymbol{z}$ represents public "bads" then $\theta < 0$ (and $\phi > 0$), the condition $0 < t < 1$ leads to an increase in utility and thus CV (EV) is equivalent to WTP (WTA) again.

The results in Theorem 1 arise when the underlying function $u(\cdot)$ is independently homogeneous. Appendix B provides results under joint homogeneity, and these additional results help provide theoretical guidance for the independent homogeneity assumption in our main results. Also, Appendix B provides results under the restrictive case where $g(\cdot)$ is the identity function leading to homogeneous utility and help provide intuition for the EV and CV closed-form solutions that we find under homotheticity in Theorem 1.

By inspection, Theorem 1 shows that CV and EV are asymmetric. Theorem 2 proves EV is greater than CV always for the same proportional change in public good provision, regardless of whether public good provision is increasing or decreasing.

**Theorem 2.** For all $t > 0$ $(t \neq 1)$, $\mathrm{EV}(m, t) > \mathrm{CV}(m, t)$. Also, the ratio $\mathrm{EV}(m, t)/\mathrm{CV}(m, t)$ increases in $\theta/\eta$ when $t > 1$, and $\mathrm{EV}(m, t)/\mathrm{CV}(m, t)$ decreases in $\theta/\eta$ when $0 < t < 1$.

Hanemann (1991) shows that EV equals CV when a public and a private good are perfect substitutes. Our assumption of independent homogeneity of the utility function in private and public good(s) rules out the possibility of perfect substitutability (see Appendix A for further discussion). Consequently, in our framework, EV always exceeds CV, but our results remain consistent with Hanemann (1991) otherwise. Regardless, in our neoclassical model, the difference between WTP and WTA for public good provision arises through two mechanisms.[7] The first mechanism comes from the budget constraint leading to a bounded WTP but an unbounded WTA. The second mechanism is the relative preference effect where the disparity between WTP and WTA is mediated by $\theta/\eta$ (as shown by Theorem 2). Therefore, a large relative preference effect can help explain the Tunçel and Hammitt (2014) finding that the WTA-WTP gap is larger for public goods than private goods.[8]

One key advantage of our framework is that it identifies the relative preference effect that drives a systematic wedge between WTP and WTA—beyond the conventional explanations like behavioral biases (e.g., loss aversion). Thus, the relative preference effect highlights that even in a fully neoclassical setting, a large discrepancy between WTP and WTA can arise if public goods are sufficiently "important" relative to the private good. When $\theta$ is large relative to $\eta$, households will display a substantial difference between what they are willing to pay for a small (or even non-marginal) increase versus what they would demand in compensation for a comparable decrease. Empirically, researchers can estimate this ratio $\theta/\eta$ from observed data and upon finding that it exceeds one, for example, indicates that WTA could be expected to surpass WTP by more than what standard consumer-surplus approximations (or purely behavioral models) might predict. This interpretation helps connect seemingly abstract parameters in our model to real-world policy scenarios—showing why certain public goods trigger unusually large WTA–WTP gaps and how one might quantify those gaps without resorting to approximate surplus measures or behavioral assumptions.

[7] In a dynamic model, Zhao and Kling (2004) finds a disparity between WTA and WTP due to uncertainty, irreversibility, and learning.

[8] Vossler *et al.* (2023) explores behavioral explanations for the gap between WTP and WTA for public goods and shows that implementing incentive-compatible mechanism to control for behavioral effects significantly closes the gap and provides valid welfare estimates.

Suppose an empirical study finds that a representative sample of households with an average annual income of $m = \$1000$ has a \$50 average willingness to pay (WTP) for a 20 percent improvement in a public good $(\boldsymbol{z})$. In our model, compensating variation is given by $\text{CV}(m, t) = \left(1 - 1.2^{\phi}\right)1000 = 50$. Solving for $(-\theta/\eta) \equiv \phi$ finds $-0.28 \approx \ln(0.95)/\ln(1.2)$ and thus implies $\theta/\eta \approx 0.28$. Interpreting this economically: a smaller $\theta/\eta$ means the public good has less impact on utility as compared to private goods, all else equal, yielding a smaller WTP for the same proportional improvement. Researchers could replicate this simple calibration approach across different sub-samples (e.g., varying incomes or demographic groups) or for different observed WTP estimates (including those from stated-preference surveys) to estimate how $\theta/\eta$ varies across the population. This showcases how one can take the theoretical closed-form solutions straight to real-world data and empirical results.

One practical application arises in urban environmental policy—for example, introduction of a new municipal water-treatment facility or a large-scale pollution-control program. Typically, local or national agencies rely on stated preference or hedonic estimates to infer how residents value cleaner water or improved air quality. Our model can be deployed in such a scenario by (i) employing existing willingness-to-pay figures from either revealed-preference (e.g., housing price differentials) or stated-preference surveys, (ii) calibrating the ratio of degrees of homotheticity $(\theta/\eta)$ to capture the public good's weight in utility relative to private goods, and then (iii) predicting how WTP or WTA scales with income or other demographic variables. This approach ensures that even non-marginal policy shifts—like a 50% reduction in air pollution such as particulate matter concentrations—can be analyzed with exact compensating or equivalent variation, rather than relying on approximate consumer-surplus measures. Local officials and researchers could thus compare policy options (e.g., partial vs. full clean-up) in terms of precise welfare benefits across different income strata and demographic characteristics, thereby linking sound theoretical underpinnings to concrete policy evaluation.

Finally, because our approach applies to non-marginal changes, it can be deployed in cases where the policy change involves a major infrastructure investment, a significant environmental improvement, or a sweeping regulatory shift. The resulting closed-form WTP and WTA formulas facilitate transparent cost–benefit analyses, especially if policymakers want to gauge winners and losers in different income brackets. By mapping real-world empirical estimates into these formulas, agencies can simulate counterfactual policies, compare expected welfare outcomes across subgroups, and design redistributive mechanisms (e.g., tiered pricing schemes or targeted transfers) to align policy with equity objectives. Thus, our framework ultimately serves as a practical toolkit that will help public agencies and researchers to translate utility-based concepts into actionable, data-driven guidance for large-scale policy decisions.

We conclude this section with Corollary 1 as it makes additional connections between the CV and EV expressions in Theorem 1 and the representative consumer's budget constraint. Then, Figure 1 provides a visual representation of Theorem 1 and Corollary 1. To start, we specify the model to have two private goods $(x_1, x_2)$, a single public good $(z)$, and let $g(\cdot)$ to be the identity function. Without loss of generality, Figure 1 identifies $\text{CV} > 0$ in the $x_1$-$x_2$ space with $t > 1$ and $\phi < 0$. Point A is the horizontal intercept of the budget constraint and equals $m/p_1$ at the original level of public good provision and income. Similarly, point B is the horizontal intercept of the budget constraint at the new level of public good provision and income, and equals $t^{\phi}\, m/p_1$. The difference between points B to A equals $\left(1 - t^{\phi}\right) m/p_1 = \text{CV}/p_1 > 0$ since $\text{CV} = \left(1 - t^{\phi}\right)m$.

Analogously, CV can be measured on the vertical $x_2$ axis too by observing the difference from points D to C equals $\left(1 - t^{\phi}\right) m/p_2 = \text{CV}/p_2 > 0$. Therefore, CV can be determined by $p_1$ times the distance between the points B and A, or, equivalently, $p_2$ times the distance between the points D and C. Thus, CV is implicitly measured twice in Figure 2, and a similar figure can be constructed to measure EV.

**Corollary 1.** Let $g(\cdot)$ be a monotonic transformation such that utility function $h(\boldsymbol{x}; \boldsymbol{z}, \boldsymbol{w}) \equiv g\big(u(\boldsymbol{x}; \boldsymbol{z}, \boldsymbol{w})\big)$ is homothetic. If the function $u(\cdot)$ is independently homogeneous (with $\eta \neq 0$) and $u(\cdot)$ is strictly quasiconcave, then CV can be expressed as follows: $\text{CV} = \sum_{n=1}^{N} \text{CV}_n$ , where $\text{CV}_n \equiv p_n(x_n^* - x_n')$ for all $n = 1, \dots, N$ for the optimal demand vectors $\boldsymbol{x}^* = \boldsymbol{x}^{\boldsymbol{h}}(u_0, \boldsymbol{p}; \boldsymbol{z}_{\boldsymbol{0}}, \boldsymbol{w})$ and $\boldsymbol{x}' = \boldsymbol{x}^{\boldsymbol{h}}(u_0, \boldsymbol{p}; t\boldsymbol{z}_{\boldsymbol{0}}, \boldsymbol{w})$, and, similarly, $\text{EV} = \sum_{n=1}^{N} \text{EV}_n$, where $\text{EV}_n \equiv p_n(x_n'' - x_n^{**})$ with $\boldsymbol{x}^{**} = \boldsymbol{x}^{\boldsymbol{h}}(u_1, \boldsymbol{p}; \boldsymbol{z}_{\boldsymbol{0}}, \boldsymbol{w})$ and $\boldsymbol{x}'' = \boldsymbol{x}^{\boldsymbol{h}}(u_1, \boldsymbol{p}; t\boldsymbol{z}_{\boldsymbol{0}}, \boldsymbol{w})$.

**Figure 1. Compensating Variation**

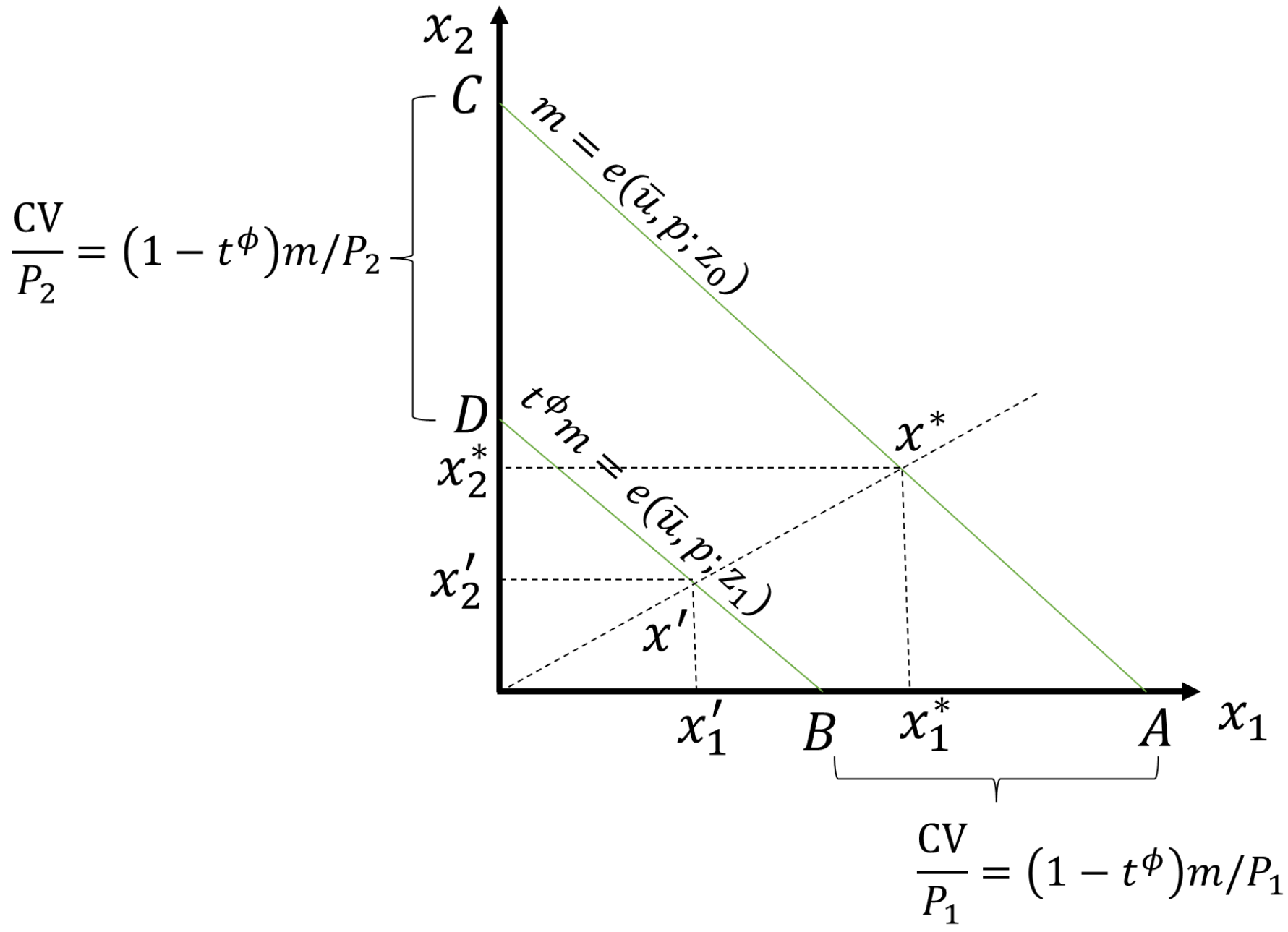

# 4 Sufficient Statistic

## 4.A Sufficient Statistic: Calculation and Applications

In this section, we calculate $\phi \equiv (-\theta/\eta)$ values using actual WTP estimates from the literature. Analogous calculations can be conducted with the EV closed-form solution and WTA estimates. Our primary aim in this section is to provide one clear, concrete example of how the closed-form formulas can be calibrated using a real-world estimate. Although we focus on Ito and Zhang (2020)'s estimate for this demonstration, we wish to emphasize that the theoretical result itself is not tied to a single setting like health or environmental economics. As explained above, the homothetic, non-separable framework can be applied to any context—public finance, industrial organization, trade, or regional economics—where researchers assume (or approximate) homothetic preferences for analytical tractability and wish to evaluate non-marginal changes. Thus, we do not regard this section as an exhaustive survey of all possible uses; rather, it illustrates how one might implement our approach in practice. Indeed, numerous other studies that rely on homothetic specifications (e.g., CES preference) could adopt our formulas with relatively minimal adaptation. Therefore, our example, although drawn from environmental valuation, is sufficiently indicative of how the framework could be applied broadly.

We begin by selecting two revealed preference studies looking at particulate matter (PM) pollution in China and compare outcomes. In the first study, Ito and Zhang (2020) finds that households in China are willing to pay \$32.7 annually to reduce PM10 concentrations from 120 micrograms per cubic meter ($\mu$g/m$^3$) to 96 $\mu$g/m$^3$, or 24 $\mu$g/m$^3$ reduction, and in our notation this means $t = 0.80$ for a reduction in a public "bad". As Section 3 discusses, applying our results to public "bad" requires $\theta < 0$ meaning $\phi > 0$. The compensating variation formula remains $\text{CV} = (1 - t^{\phi})m$, but $\text{CV} > 0$ if $0 < t < 1$. Also, rearranging the compensating variation formula from Theorem 1 yields $\phi = \ln\left(1 - \frac{CV}{m}\right)/\ln(t)$. The average annual income of the affected households in Ito and Zhang (2020)'s study is \$527.5/year and thus $\phi = 0.29$. In the second study, Freeman *et al.* (2019) finds a WTP of \$316.4 annually for a one standard deviation decrease in PM2.5, equal to 14.6 $\mu\text{g/m}^3$ from a baseline of 41.4 $\mu\text{g/m}^3$, and implying $t = 0.65$.

Freeman *et al.* (2019) also studies Chinese households, but the sample population is richer with a median income of \$1481.2/year. Again, using the formula in Theorem 1 and solving then yields $\phi = 0.55$.

Next, we use our CV formula to provide plausible estimates of WTP for similar populations in the same setting; specifically, we show how to find WTP estimates for different income groups. Then, we compare our estimates to those in the literature that estimate WTP heterogeneity by income. Our case study to demonstrate these methods uses the results from Ito and Zhang (2020). Their study employs a rich dataset and a computationally challenging, random-coefficient logit model to uncover WTP heterogeneity across household income. Ito and Zhang (2020) reports that the 95-percent confidence interval for WTP is approximately \$12 to \$36 annually for households with income of \$1000/year using their random-coefficient estimates, where \$1000/year is more the two standard deviations above the average household income in their sample.[9] Importantly, Ito and Zhang (2020) finds an approximately linear relationship between income and WTP for the vast majority of the income distribution, only deviating slightly at the upper tail of the distribution, and this linearity matches the linear relationship in the Theorem 1's closed-form solutions.

---

[9] The average income in their sample is \$527.5/year with a standard deviation of \$153, and the WTP for PM10 reductions overall is \$32.7/year. This suggests a skewed income distribution in their sample as WTP estimates by income in Ito and Zhang (2020) go up to \$10,000/year (or more than 68 standard deviations above the mean). See Ito and Zhang (2020)'s Figure 4.

We use Theorem 1's CV expression along with Ito and Zhang (2020)'s simpler standard logit WTP estimate to calculate WTP values for other income groups without needing to deploy a random-coefficient model. Specifically, the Ito and Zhang (2020) standard logit estimate for the marginal WTP (MWTP) is approximately \$1.25 per $\mu\text{g/m}^3$ across all income groups. Thus, a 24 $\mu\text{g/m}^3$ reduction yields an approximate total WTP of \$30 annually. This implies $\phi = 0.26$ at the average income for $t = 0.80$. Then, applying the CV expression with the new $\phi$, but with an income of \$1000/year, finds a WTP of \$56.0 annually. This WTP approximation is slightly larger than the upper bound of Ito and Zhang (2020)'s random-coefficient model confidence interval that is centered on \$36/year. Regardless, the estimate of \$56/year could feasibly be used in policy analysis despite \$1000/year being far from the average household income. Finally, we have not yet considered errors in the estimation of pollution reduction (or any other estimate parameters). Indeed, only a slight change in the pollution reduction ratio to $t = 0.87$ recovers a WTP of \$36/year, and $t = 0.87$ is with is within the range implied by Ito and Zhang (2020)'s 95-percent confidence interval for pollution reduction.[10]

## 4.B Sufficient Statistic: Identification and Estimation

If an empirical study is unavailable, then the below methodology shows how to identify and estimate the sufficient statistic from data directly. Again, the analysis and discussion below in this section addresses CV and WTP directly, but the methods can be easily adapted for our closed-form EV expression and thus WTA. Theorem 1 provides the equality $t^{-\theta/\eta}e(u,\boldsymbol{p};\boldsymbol{z},\boldsymbol{w}) = e(u,\boldsymbol{p};t\boldsymbol{z},\boldsymbol{w})$, so taking the logarithm of both sides and rearranging yields:

$$\ln\left[\frac{m_1}{m_0}\right] = (-\theta/\eta)\ln[t] \quad (1)$$

with $\boldsymbol{z_1} = t\boldsymbol{z}_0$ and $m_0 = e(u,\boldsymbol{p};\boldsymbol{z_0},\boldsymbol{w})$, and also let $m_1 = e(u,\boldsymbol{p};t\boldsymbol{z_0},\boldsymbol{w}) = e(u,\boldsymbol{p};\boldsymbol{z_1},\boldsymbol{w})$ be the income after the change in the public good provision that maintains a constant level of utility. Then, define the change public good provision as $\Delta z \equiv \|\boldsymbol{z}_1\|/\|\boldsymbol{z}_0\| = t$, and similarly define $\Delta m \equiv m_1/m_0$.

[10] Ito and Zhang (2020)'s main estimate for pollution reduction is 24.54 $\mu\text{g/m}^3$ with a standard error of 6.97 $\mu\text{g/m}^3$. From a baseline of 120 $\mu\text{g/m}^3$, this implies calculated value of $t$ then ranges from 0.68 to 0.91 such that 0.87 falls within that range.

Next, we translate equation (1) to an estimating equation and start by defining the following variables: $Y$ = log of the change in expenditure or income = $\ln[\Delta m]$; and, $X_1$ = log of the change in public good provision = $\ln[\Delta z]$. Then, using equation (1), we specify an estimating equation:

$$Y = \beta_0 + \beta_1 X_1 + \varepsilon \quad (2)$$

where $\varepsilon$ is the error term.[11] The parameter of interest is $\beta_1$ equal to $\phi \equiv (-\theta/\eta)$, and $\phi$ can be estimated using quasi-experiment techniques such as instrumental variables or regression discontinuity. The data needed to estimate equation (2) can be individual-level observations or aggregate-level data depending on the setting. For example, equation (2) allows the use of aggregate data such as national income to account for $m$ since the structural assumption of homogeneity enable aggregation over individuals. However, income data subject to the requirements of quasi-experimental causal identification may be unavailable, so we next turn to identifying and estimating $\phi$ using consumption data.

---

[11] The upper-case $X_i$ variables in the estimating equation should not be confused with the lower-case $x_i$ values for private goods in our utility model.

Alternatively, although under the stronger assumption of strictly quasiconcave utility, Theorem 1 finds $t^{-\theta/\eta}x_n^h(u,\boldsymbol{p};\boldsymbol{z},\boldsymbol{w}) = x_n^h(u,\boldsymbol{p};t\boldsymbol{z},\boldsymbol{w})$ for all $n = 1,\dots,N$ meaning that the equality holds for all private goods. Again, taking the logarithm of both sides and rearranging yields:

$$\ln\left[\frac{x_{n1}}{x_{n0}}\right] = (-\theta/\eta)\ln[t] \;\forall\, n = 1,\dots,N \tag{3}$$

where $x_{n0} = x_n^h(u,\boldsymbol{p};\boldsymbol{z}_0,\boldsymbol{w})$ and $x_{n1} = x_n^h(u,\boldsymbol{p};t\boldsymbol{z}_0,\boldsymbol{w}) = x_n^h(u,\boldsymbol{p};\boldsymbol{z}_1,\boldsymbol{w})$ are the private consumption quantities related to the level of public good provision. Then, define the change in private good consumption as $\Delta x_n \equiv x_{n1}/x_{n0}$, and redefine the dependent variable as $Y'$ = log of the change in private consumption of the $n^{\text{th}}$ good = $\ln[\Delta x_n]$. Thus, we can specify an alternative estimating equation:

$$Y_n' = \beta_{0n} + \beta_{1n}X_1 + \varepsilon_n' \;\forall\, n = 1,\dots,N \tag{4}$$

where $\varepsilon_n'$ is the error term for each good and noting $X_1 = \ln[t]$ is common across private goods. Then, to identify $\beta_{1n}$, equation (4) can be estimated for a single good using quasi-experimental methods in a similar manner to equation (2). However, if data on multiple private goods is available at the individual- or aggregate-level, then (4) becomes a system of seemingly unrelated regression equations to estimate the parameter of interest. Furthermore, *ex-post* inference can check if $\beta_{1n}$ is the same across all equations or the researcher can constrain $\beta_{1n}$ to be the same across equations. As we discuss above, Ito and Zhang (2020) provides a good example where data could be used to calculate $\phi$ directly as that study employs a regression discontinuity method to find the change in private air filter purchases $(x_n)$ for a change in particulate matter pollution $(t)$.

The goal when estimating equation (2) or (4) is to identify and estimate the parameter $\beta_1 = \phi \equiv (-\theta/\eta)$ noting that the same $X_1$ appears in both equations. Importantly, it is not necessary to separately identify and estimate the parameters $\eta$ and $\theta$. Rather, it is only necessary to identify and estimate the ratio $-\theta/\eta$ as this ratio appears in the CV expression (and similarly for EV). That is, $\phi$ is a sufficient statistic in the statistical meaning (Reid, 2015), and, relatedly, infinite combinations of the primitives $\eta$ and $\theta$ generate the same $\phi$. Also, the sign of $\phi$ determines the sign of CV, so $\phi$ is a sufficient statistic in the economic meaning as well (Chetty, 2009). The level values of the preference parameters $\eta$ and $\theta$, all else equal, do not affect the CV expression because the underlying result in Theorem 1 only uses the ordinal properties of the homogeneous function $u(\boldsymbol{x}; \boldsymbol{z}, \boldsymbol{w})$ that translate to the homothetic utility function $h(\boldsymbol{x}; \boldsymbol{z}, \boldsymbol{w})$. It is the relative importance of the public goods to the private goods as expressed by the ratio $\theta/\eta$ that matters.

We note that $\ln[t]$ is needed in both estimating equations where $t > 0$ is the scaling of public good provision in either quality and/or the quantity of the public goods. The vector $\boldsymbol{z}$ can be a single, scalar public good, a vector of public goods summarized by a single statistic ($z = g(\boldsymbol{z})$), or a vector of public goods that all change by the same quantity and/or quality. The variable definitions prior to equation (4) in this section say "expenditure or income" due to the single-period utility maximization model of the representative consumer.[12]

# 5 Additively Separable Utility

In this last section, we address the special case of additively separable utility since this assumption often explicitly or implicitly arises in many theoretical and empirical contexts. We will focus on utility functions with only $\boldsymbol{z}$. That is, we drop $\boldsymbol{w}$ from the model. Under additivity, the utility function can be written as $u(\boldsymbol{x};\boldsymbol{z}) = u_1(\boldsymbol{x}) + u_2(\boldsymbol{z})$ where $u_2(\cdot)$ is an additive utility term containing the public goods. That is, public goods influence only the level of utility but not consumer choices on the margin; for instance, if more public parks make a consumer better off but do not change the marginal incentive to buy hiking boots. However, Example 1 shows that a utility function with additively separable public goods does not necessarily lead to a homogeneous expenditure function in public goods and thus Theorem 1 does not hold. Rather, Example 1 suggests that the expenditure function under additive separability includes the term "$u - u_2(\boldsymbol{z})$" and Proposition 1 proves that this is true for all additively separable utility functions, and vice versa.

[12] Diewart (2012) provides necessary and sufficient conditions using Afriat's Theorem to check whether finite price and quantity data are consistent with homothetic utility maximization.

**Example 1:** Let $u(\boldsymbol{x};\boldsymbol{z}) = u_1(\boldsymbol{x}) + u_2(\boldsymbol{z}) = (x_1x_2)^{1/\alpha} + (z_{11}z_{12})^{1/\beta}$ with $\boldsymbol{z} = (z_{11}\ z_{12})$ where $u(\boldsymbol{x};\boldsymbol{z})$ is jointly homogeneous of degree $\gamma = 2/\delta$ when $\delta = \alpha = \beta$. Assume the function is jointly homogeneous and then expenditure function is found to be $e(u,\boldsymbol{p};\boldsymbol{z}) = 2\left(u - (z_{11}z_{12})^{1/\delta}\right)^{\delta/2}(p_1p_2)^{1/2}$. However, the expenditure function is not homogeneous in the public goods as shown by $e(u,\boldsymbol{p};t\boldsymbol{z}) = 2\left(u - (z_{11}z_{12})^{1/\delta}(t)^{2/\delta}\right)^{\delta/2}(p_1p_2)^{1/2}$.

**Proposition 1.** The utility function is additively separable in the public good, such that $u(\boldsymbol{x};\boldsymbol{z}) = u_1(\boldsymbol{x}) + u_2(\boldsymbol{z})$, if and only if the expenditure function can be expressed as $e(u,\boldsymbol{p};\boldsymbol{z}) = e(u - u_2(\boldsymbol{z}),\boldsymbol{p})$.

Finally, Proposition 2 finds an expression for the expenditure function in terms of indirect utility and the degree of homogeneity of $u_1(\boldsymbol{x})$. This result explicitly relates private expenditure to public goods despite additive separability. However, as Example 1 demonstrates, it does not allow for the compensating or equivalent variation expressions for public goods considered by our main theorem in Section 3.

**Proposition 2.** If a utility function is additively separable in public good and $u_1(\boldsymbol{x})$ is homogeneous of degree $\gamma \neq 0$, then the expenditure function has the form $e(u, \boldsymbol{p}; \boldsymbol{z}) = \left[\frac{u - u_2(\boldsymbol{z})}{v_1(1, \boldsymbol{p})}\right]^{1/\gamma}$.

# 6 Conclusion

Governments incur billions of dollars in costs to provide public goods but the willingness to pay (WTP) and willingness to accept (WTA) for changes in public goods are not directly observable by policymakers. This study derives exact closed-form compensating variation (CV) and equivalent variation (EV) expressions for a change in public good provision, where the change can be in the quantity and/or quality of the public good(s). These closed-form expressions provide a concrete bridge between the theorems in this study and a wide range of real-world applications across fields and settings that employ WTP and WTA for public goods.

The CV and EV expressions we find are linear in initial income but mediated by preference parameters from a homothetic utility function with an independently homogeneous underlying function. We also show two different ways to estimate a single sufficient statistic for the two preference parameters in our expressions. Furthermore, we identify three mechanisms – that we call the income effect, absolute preference effect, and relative preference effect – that drive CV and EV in our model and help explain the gap between WTP and WTA for public goods. Our results contribute to the literature by providing exact CV and EV expressions for both marginal and non-marginal changes in public good provision for a family of homothetic utility functions. In addition, we show how to derive the sufficient statistic from WTP (WTA) estimates in the literature and then how those estimates can be used to calculate WTP (WTA) across income groups.

The model generating our CV and EV expressions are for homothetic utility functions that encompass a wide spectrum of functions that have been employed in contemporary theoretical and empirical economic studies. The CV and EV expressions derived here remain valid when an auxiliary public goods vector is included in the utility function but not subject to the homogeneity condition; rather, only the public good of interest must manifest independent homogeneity. Furthermore, the much-employed Willig condition also has a restriction with respect to income (Smith and Banzhaf, 2004) and employs consumer surplus as an approximate welfare measure because demand is not directly observed for public goods (Bockstael and McConnell, 1993). This limits the conventional willingness-to-pay analysis to marginal changes in public goods, but our closed-form solutions are not limited to marginal changes. In addition, our CV and EV expressions do not use consumer surplus as a welfare approximation and thus provides exact WTP and WTA measures for both marginal and non-marginal changes in public good provision.

# Appendix A

**Proof of Theorem 1.** Part (a): Begin with the EMP definition:

$e(u, \boldsymbol{p}; \boldsymbol{z}, \boldsymbol{w}) = \min_{\boldsymbol{x}}\{\boldsymbol{p} \cdot \boldsymbol{x}: h(\boldsymbol{x}; \boldsymbol{z}, \boldsymbol{w}) \geq u\}$

$\Leftrightarrow e(u, \boldsymbol{p}; \boldsymbol{z}, \boldsymbol{w}) = \min_{\boldsymbol{x}}\{\boldsymbol{p} \cdot \boldsymbol{x}: g\big(u(\boldsymbol{x}; \boldsymbol{z}, \boldsymbol{w})\big) \geq u\}$ [Definition]

$\Leftrightarrow e(u, \boldsymbol{p}; t\boldsymbol{z}, \boldsymbol{w}) = \min_{\boldsymbol{x}}\{\boldsymbol{p} \cdot \boldsymbol{x}: g\big(u(\boldsymbol{x}; t\boldsymbol{z}, \boldsymbol{w})\big) \geq u\}$

$\Leftrightarrow e(u, \boldsymbol{p}; t\boldsymbol{z}, \boldsymbol{w}) = \min_{\boldsymbol{x}}\left\{\boldsymbol{p} \cdot \boldsymbol{x}: g\left(t^{\theta} u(\boldsymbol{x}; \boldsymbol{z}, \boldsymbol{w})\right) \geq u\right\}$ [Assumption: Homogeneity of $u(\cdot)$ in $\boldsymbol{z}$]

$\Leftrightarrow e(u, \boldsymbol{p}; t\boldsymbol{z}, \boldsymbol{w}) = \min_{\boldsymbol{x}}\left\{\boldsymbol{p} \cdot \boldsymbol{x}: g\left(t^{\theta} u\left(\frac{t^{(-\theta/\eta)}}{t^{(-\theta/\eta)}}\boldsymbol{x}; \boldsymbol{z}, \boldsymbol{w}\right)\right) \geq u\right\}$

$\Leftrightarrow e(u, \boldsymbol{p}; t\boldsymbol{z}, \boldsymbol{w}) = \min_{\boldsymbol{x}}\left\{\boldsymbol{p} \cdot \boldsymbol{x}: g\left(t^{\theta} t^{((-\theta/\eta)\eta)} u\left(\frac{1}{t^{(-\theta/\eta)}}\boldsymbol{x}; \boldsymbol{z}, \boldsymbol{w}\right)\right) \geq u\right\}$ [Assumption: Homogeneity of $u(\cdot)$ in $\boldsymbol{x}$]

$\Leftrightarrow e(u, \boldsymbol{p}; t\boldsymbol{z}, \boldsymbol{w}) = \min_{\boldsymbol{x}}\left\{\boldsymbol{p} \cdot \boldsymbol{x}: g\left(u\left(\frac{1}{t^{(-\theta/\eta)}}\boldsymbol{x}; \boldsymbol{z}, \boldsymbol{w}\right)\right) \geq u\right\}$

$\Leftrightarrow e(u, \boldsymbol{p}; t\boldsymbol{z}, \boldsymbol{w}) = \min_{\widetilde{\boldsymbol{x}}}\{\boldsymbol{p} \cdot t^{(-\theta/\eta)}\widetilde{\boldsymbol{x}}: g(u(\widetilde{\boldsymbol{x}}; \boldsymbol{z}, \boldsymbol{w}) \geq u)\}$ [Unit conversion: $\widetilde{\boldsymbol{x}} = \frac{1}{t^{(-\theta/\eta)}}\boldsymbol{x}$]

$\Leftrightarrow e(u, \boldsymbol{p}; t\boldsymbol{z}, \boldsymbol{w}) = t^{(-\theta/\eta)} \min_{\widetilde{\boldsymbol{x}}}\{\boldsymbol{p} \cdot \widetilde{\boldsymbol{x}}: g\big(u(\widetilde{\boldsymbol{x}}; \boldsymbol{z}, \boldsymbol{w})\big) \geq u\}$ [Min operator property]

$\Leftrightarrow e(u, \boldsymbol{p}; t\boldsymbol{z}, \boldsymbol{w}) = t^{(-\theta/\eta)} \min_{\widetilde{\boldsymbol{x}}}\{\boldsymbol{p} \cdot \widetilde{\boldsymbol{x}}: h(\widetilde{\boldsymbol{x}}; \boldsymbol{z}, \boldsymbol{w}) \geq u\}$ [Definition]

$\Leftrightarrow e(u, \boldsymbol{p}; t\boldsymbol{z}, \boldsymbol{w}) = t^{(-\theta/\eta)} e(u, \boldsymbol{p}; \boldsymbol{z}, \boldsymbol{w})$ [Definition].

Part (b): When $u(\boldsymbol{x}; \boldsymbol{z}, \boldsymbol{w})$ is independently homogeneous then from part (a) we know $t^{-\theta/\eta} e(u, \boldsymbol{p}; \boldsymbol{z}, \boldsymbol{w}) = e(u, \boldsymbol{p}; t\boldsymbol{z}, \boldsymbol{w})$. Let the Hicksian demands in vector form be defined as $\boldsymbol{x}^* \equiv \boldsymbol{x}^{\boldsymbol{h}}(u, \boldsymbol{p}; \boldsymbol{z}, \boldsymbol{w})$ and $\boldsymbol{x}' \equiv \boldsymbol{x}^{\boldsymbol{h}}(u, \boldsymbol{p}; t\boldsymbol{z}, \boldsymbol{w})$. Since the utility level is fixed, then it follows:

$h(\boldsymbol{x}^*; \boldsymbol{z}, \boldsymbol{w}) = h(\boldsymbol{x}'; t\boldsymbol{z}, \boldsymbol{w})$

$\Leftrightarrow g\big(u(\boldsymbol{x}^*; \boldsymbol{z}, \boldsymbol{w})\big) = g\big(u(\boldsymbol{x}'; t\boldsymbol{z}, \boldsymbol{w})\big)$

$\Leftrightarrow u(\boldsymbol{x}^*; \boldsymbol{z}, \boldsymbol{w}) = u(\boldsymbol{x}'; t\boldsymbol{z}, \boldsymbol{w})$ [Apply $g^{-1}(\cdot)$ to both sides]

$\Leftrightarrow u(\boldsymbol{x}^*; \boldsymbol{z}, \boldsymbol{w}) = t^{\theta} u(\boldsymbol{x}'; \boldsymbol{z}, \boldsymbol{w})$ [Assumption: Homogeneity of $u(\cdot)$ in $\boldsymbol{z}$]

$\Leftrightarrow t^{-\theta} u(\boldsymbol{x}^*; \boldsymbol{z}, \boldsymbol{w}) = u(\boldsymbol{x}'; \boldsymbol{z}, \boldsymbol{w})$

$\Leftrightarrow u\left(t^{-\theta/\eta} \boldsymbol{x}^*; \boldsymbol{z}, \boldsymbol{w}\right) = u(\boldsymbol{x}'; \boldsymbol{z}, \boldsymbol{w})$ [Assumption: Homogeneity of $u(\cdot)$ in $\boldsymbol{x}$]

and therefore $t^{-\theta/\eta} \boldsymbol{x}^* = \boldsymbol{x}'$ by strict quasi-concavity leading to uniqueness. Alternatively, assuming a differentiable expenditure function, this part of the corollary can be proven by differentiating the expenditure function and applying Shepard's Lemma.

Part (c): To start, define the utility levels given different levels of public goods provision as follows:

$$u_0 \equiv v(m, \boldsymbol{p}; \boldsymbol{z}_0, \mathbf{w})$$

$$u_1 \equiv v(m, \boldsymbol{p}; \boldsymbol{z}_1, \mathbf{w})$$

with $\boldsymbol{z}_1 \equiv t\boldsymbol{z}_0$ (for $t > 0$) where $\boldsymbol{z}_0, \boldsymbol{z}_1 \in \boldsymbol{z}$. We call $\boldsymbol{z}_0$ the "initial" of "original" level of public good provision and $\boldsymbol{z}_1$ the "new" or "scaled" level of public good provision. Recall that Mäler's indirect utility definition for CV found in Hanemann (1991) is given as:

$$v(\mathrm{m} - \mathrm{CV}, \boldsymbol{p}; \boldsymbol{z}_1, \mathbf{w}) = v(m, \boldsymbol{p}; \boldsymbol{z}_0, \mathbf{w}) = u_0$$

where $m$ is the fixed income level from the utility maximization problem (UMP). For the CV expression, $m$ and $m - \mathrm{CV}$ are the levels of expenditure needed to reach utility level $u_0$ with public good levels $\boldsymbol{z}_0$ and $\boldsymbol{z}_1$, respectively. In turn, this means:

$$m = e(u_0, \boldsymbol{p}; \boldsymbol{z}_0, \mathbf{w})$$

$$m - \mathrm{CV} = e(u_0, \boldsymbol{p}; \boldsymbol{z}_1, \mathbf{w})$$

and subtracting the equations finds:

$$\mathrm{CV} = e(u_0, \boldsymbol{p}; \boldsymbol{z}_0, \mathbf{w}) - e(u_0, \boldsymbol{p}; \boldsymbol{z}_1, \mathbf{w}) \Leftrightarrow \mathrm{CV} = e(u_0, \boldsymbol{p}; \boldsymbol{z}_0, \mathbf{w}) - e(u_0, \boldsymbol{p}; t\boldsymbol{z}_0, \mathbf{w}).$$

If $t > 1$, then $\boldsymbol{z}_1 > \boldsymbol{z}_0$ meaning $e(u_0, \boldsymbol{p}; \boldsymbol{z}_0, \boldsymbol{w}) > e(u_0, \boldsymbol{p}; \boldsymbol{z}_1, \boldsymbol{w})$, implying $\text{CV} > 0$, since a higher level of public good provision requires a lower level of expenditure to achieve utility $u_0$. For this increase in public good provision, $\text{CV} > 0$ is an individual's willingness to pay (WTP) to go from an initial, lower $(\boldsymbol{z}_0)$ to a new, higher $(\boldsymbol{z}_1)$ level of public good provision.

In contrast, if $0 < t < 1$, then $\boldsymbol{z}_0 > \boldsymbol{z}_1$ and $\text{CV} < 0$. Here, compensating variation measures the willingness to accept (WTA) a decrease in public good provision from the initial to the new level. As such, the absolute value of CV is the amount that an individual must be paid to return to the initial utility $u_0$.

Similarly, Mäler's indirect utility definition for EV found in Hanemann (1991) is given as:

$$v(m, \boldsymbol{p}; \boldsymbol{z}_1, \mathbf{w}) = v(m + \text{EV}, \boldsymbol{p}; \boldsymbol{z}_0, \mathbf{w}) = u_1.$$

For the EV expression, $m$ and $m + \text{EV}$ are the levels of expenditure needed to reach utility level $u_1$ with public good levels $\boldsymbol{z}_1$ and $\boldsymbol{z}_0$, respectively. In turn, this means:

$$m = e(u_1, \boldsymbol{p}; \boldsymbol{z}_1, \mathbf{w})$$

$$m + \text{EV} = e(u_1, \boldsymbol{p}; \boldsymbol{z}_0, \mathbf{w})$$

and subtracting the equations finds:

$$-\text{EV} = e(u_1, \boldsymbol{p}; \boldsymbol{z}_1, \mathbf{w}) - e(u_1, \boldsymbol{p}; \boldsymbol{z}_0, \mathbf{w})$$

$$\Leftrightarrow \text{EV} = e(u_1, \boldsymbol{p}; \boldsymbol{z}_0, \mathbf{w}) - e(u_1, \boldsymbol{p}; \boldsymbol{z}_1, \mathbf{w}) \Leftrightarrow \text{EV} = e(u_1, \boldsymbol{p}; \boldsymbol{z}_{\mathbf{0}}, \mathbf{w}) - e(u_1, \boldsymbol{p}; \boldsymbol{t}\boldsymbol{z}_0, \mathbf{w}).$$

If $t > 1$, then $\boldsymbol{z}_1 > \boldsymbol{z}_0$ meaning $e(u_1, \boldsymbol{p}; \boldsymbol{z}_0, \mathbf{w}) - e(u_1, \boldsymbol{p}; \boldsymbol{z}_1, \mathbf{w})$, implying $\text{EV} > 0$, since EV is the amount that an individual must be paid (i.e., willingness to accept) when returning to the original level of public good provision $\boldsymbol{z}_0$ but still maintaining the new level of utility $u_1$. In other words, EV is the amount an individual would accept to forgo the higher level of public good provision while keeping to the same level of utility. If $0 < t < 1$, then $\text{EV} < 0$ represents the amount an individual would be willing to pay to return to $u_1$ since the new level of public good provision $\boldsymbol{z}_1$ is less than the original level $\boldsymbol{z}_0$.

We continue part (c) of the proof as follows. For CV,

$\mathrm{CV}(\boldsymbol{z}_0, t) = e(u_0, \boldsymbol{p}; \boldsymbol{z}_0, \mathbf{w}) - e(u_0, \boldsymbol{p}; t\boldsymbol{z}_0, \mathbf{w})$ [Definition]

$\Leftrightarrow \mathrm{CV}(\boldsymbol{z}_0, t) = e(u_0, \boldsymbol{p}; \boldsymbol{z}_0, \boldsymbol{w}) - t^{\phi} e(u_0, \boldsymbol{p}; \boldsymbol{z}_0, \boldsymbol{w})$ [Theorem 1 part (a)]

$\Leftrightarrow \mathrm{CV}(\boldsymbol{z}_0, t) = \left(1 - t^{\phi}\right) e(u_0, \boldsymbol{p}; \boldsymbol{z}_0, \boldsymbol{w})$

$\Leftrightarrow \mathrm{CV}(m, t) = \left(1 - t^{\phi}\right) m.$

Alternatively,

$\mathrm{CV}(\boldsymbol{z}_0, t) = e(u_0, \boldsymbol{p}; \boldsymbol{z}_0, \mathbf{w}) - e(u_0, \boldsymbol{p}; t\boldsymbol{z}_0, \mathbf{w})$ [Definition]

$\Leftrightarrow \mathrm{CV}(\boldsymbol{z}_0, t) = e(u_0, \boldsymbol{p}; t^{-1}\boldsymbol{z}_0, \mathbf{w}) - e(u_0, \boldsymbol{p}; \boldsymbol{z}_1, \mathbf{w})$ [since $\boldsymbol{z}_1 \equiv t\boldsymbol{z}_0$]

$\Leftrightarrow \mathrm{CV}(\boldsymbol{z}_0, t) = t^{-\phi} e(u_0, \boldsymbol{p}; \boldsymbol{z}_0, \mathbf{w}) - e(u_0, \boldsymbol{p}; \boldsymbol{z}_1, \mathbf{w})$ [Theorem 1 part (a)]

$\Leftrightarrow \mathrm{CV}(\boldsymbol{z}_0, t) = \left(t^{-\phi} - 1\right) e(u_0, \boldsymbol{p}; \boldsymbol{z}_1, \mathbf{w})$

$\Leftrightarrow \mathrm{CV} = \left(t^{-\phi} - 1\right)(m - \mathrm{CV})$ [Definition]

$\Leftrightarrow \mathrm{CV} = t^{-\phi}(m - \mathrm{CV}) - (m - \mathrm{CV}) \Leftrightarrow \mathrm{CV} = t^{-\phi}(m - \mathrm{CV}) - m + \mathrm{CV}$

$\Leftrightarrow 0 = t^{-\phi}(m - \mathrm{CV}) - m \Leftrightarrow 0 = m - \mathrm{CV} - t^{\phi} m \Leftrightarrow \mathrm{CV} = m - t^{\phi} m$

$\Leftrightarrow \mathrm{CV}(m, t) = \left(1 - t^{\phi}\right) m.$

Similarly, for EV, we have:

$\mathrm{EV}(\boldsymbol{z}_0, t) = e(u_1, \boldsymbol{p}; \boldsymbol{z}_0, \boldsymbol{w}) - e(u_1, \boldsymbol{p}; t\boldsymbol{z}_0, \boldsymbol{w})$ [Definition]

$\Leftrightarrow \mathrm{EV}(\boldsymbol{z}_1, t) = e(u_1, \boldsymbol{p}; t^{-1}\boldsymbol{z}_1, \boldsymbol{w}) - e(u_1, \boldsymbol{p}; \boldsymbol{z}_1, \boldsymbol{w})$ [since $\boldsymbol{z}_1 \equiv t\boldsymbol{z}_0$]

$\Leftrightarrow \mathrm{EV}(\boldsymbol{z}_1, t) = t^{-\phi} e(u_1, \boldsymbol{p}; \boldsymbol{z}_1, \boldsymbol{w}) - e(u_1, \boldsymbol{p}; \boldsymbol{z}_1, \boldsymbol{w})$ [Theorem 1 part (a)]

$\Leftrightarrow \mathrm{EV}(\boldsymbol{z}_1, t) = t^{-\phi} e(u_1, \boldsymbol{p}; \boldsymbol{z}_1, \boldsymbol{w}) - e(u_1, \boldsymbol{p}; \boldsymbol{z}_1, \boldsymbol{w})$

$\Leftrightarrow \text{EV}(\boldsymbol{z}_1, t) = \left(t^{-\phi} - 1\right) e(u_1, \boldsymbol{p}; \boldsymbol{z}_1, \boldsymbol{w})$

$\Leftrightarrow \text{EV}(m, t) = \left(t^{-\phi} - 1\right) m.$

Alternatively,

$\text{EV}(\boldsymbol{z}_0, t) = e(u_1, \boldsymbol{p}; \boldsymbol{z_0}, \boldsymbol{w}) - e(u_1, \boldsymbol{p}; t\boldsymbol{z}_0, \boldsymbol{w})$ [Definition]

$\Leftrightarrow \text{EV}(\boldsymbol{z}_0, t) = e(u_1, \boldsymbol{p}; \boldsymbol{z_0}, \boldsymbol{w}) - t^{\phi} e(u_1, \boldsymbol{p}; t\boldsymbol{z}_0, \boldsymbol{w})$ [Theorem 1 part (a)]

$\Leftrightarrow \text{EV}(\boldsymbol{z}_0, t) = \left(1 - t^{\phi}\right) e(u_1, \boldsymbol{p}; \boldsymbol{z}_0, \boldsymbol{w})$

$\Leftrightarrow \text{EV} = \left(1 - t^{\phi}\right)(m + \text{EV})$ [Definition]

$\Leftrightarrow \text{EV} = (m + \text{EV}) - t^{\phi}(m + \text{EV}) \Leftrightarrow 0 = m - t^{\phi}(m + \text{EV}) \Leftrightarrow 0 = t^{-\phi} m - (m + \text{EV})$

$\Leftrightarrow 0 = t^{-\phi} m - (m + \text{EV}) \Leftrightarrow 0 = t^{-\phi} m - m - \text{EV} \Leftrightarrow \text{EV} = t^{-\phi} m - m$

$\Leftrightarrow \text{EV}(m, t) = \left(t^{-\phi} - 1\right) m.$ *Q.E.D.*

**Proof of Theorem 2.**

Define $\rho \equiv \theta/\eta$. Begin with $\text{EV}(m, t) > \text{CV}(m, t)$ for a fixed $t > 0$, and proceed as follows:

$\left(t^{\theta/\eta} - 1\right) m > \left(1 - t^{-\theta/\eta}\right) m$ [Theorem 1]

$\Leftrightarrow (t^{\rho} - 1) m = \text{EV}(\rho) > \text{CV}(\rho) = (1 - t^{-\rho}) m$ [Define EV and CV as functions of $\rho$]

$\Leftrightarrow t^{\rho} + t^{-\rho} - 2 > 0$

$\Leftrightarrow \left(t^{\rho/2} - t^{-\rho/2}\right)^2 > 0$ and this last statement is true for all $\rho$ and $t$, thus EV > CV, always.

Next, calculate the derivative of $\text{EV}(\rho)/\text{CV}(\rho)$. To start, $\frac{dEV}{d\rho} = mt^{\rho} \ln t$ and $\frac{dCV}{d\rho} = mt^{-\rho} \ln t$.

Applying the quotient rule finds:

$$\frac{d}{d\rho}\left[\frac{\text{EV}(\rho)}{\text{CV}(\rho)}\right] = \frac{d\text{EV}(\rho)\text{CV}(\rho) - \text{EV}(\rho)\text{dCV}(\rho)}{[\text{CV}(\rho)]^2}$$

$$\Leftrightarrow \frac{(mt^{\rho}\ln t)(1-t^{-\rho})m-(t^{\rho}-1)m(mt^{-\rho}\ln t)}{[(1-t^{-\rho})m]^2}$$

$$\Leftrightarrow \frac{(t^{\rho}\ln t)(1-t^{-\rho})-(t^{\rho}-1)(t^{-\rho}\ln t)}{(1-t^{-\rho})^2}$$

$$\Leftrightarrow \ln t\cdot\frac{(t^{\rho})(1-t^{-\rho})-(t^{\rho}-1)(t^{-\rho})}{(1-t^{-\rho})^2}$$

$$\Leftrightarrow \ln t\cdot\frac{(t^{\rho})(1-t^{-\rho})-(1-t^{-\rho})}{(1-t^{-\rho})^2}$$

$$\Leftrightarrow \ln t\cdot\left(\frac{t^{\rho}-1}{1-t^{-\rho}}\right)=\ln t\cdot\left(\frac{(t^{\rho}-1)m}{(1-t^{-\rho})m}\right)=\ln t\cdot\left(\frac{\mathrm{EV}}{\mathrm{CV}}\right).$$

For public goods, $\theta > 0$ and $\rho > 0$, and increasing a public good implies $t > 1$. As shown above, $\mathrm{EV} > \mathrm{CV} > 0$ and $\ln t > 0$ for $t > 1$, so the ratio increases. For $0 < t < 1$, then $\mathrm{CV} < \mathrm{EV} < 0$, but the ratio $\frac{\mathrm{EV}}{\mathrm{CV}} > 0$ and $\ln t < 0$, so the ratio decreases. *Q.E.D.*

The inequality $\mathrm{EV} > \mathrm{CV}$ is a consequence of private and the public goods being (imperfect) substitutes under the independent homogeneity assumption. To begin, the inequality $\mathrm{EV} > \mathrm{CV}$ is equivalent to:

$$e(u_1,p,z_0,w)-e(u_1,p,tz_0,w) > e(u_0,p,z_0,w)-e(u_0,p,tz_0,w)$$

and this can be rewritten as:

$$e(u_1,p,z_0,w)-e(u_0,p,z_0,w) > e(u_1,p,tz_0,w)-e(u_0,p,tz_0,w). \qquad \text{(A1)}$$

Consider a small increase in the utility around $u_0$ to $u_1$. Then, the first-order Taylor expansions of the expenditure function in utility are given:

$$e(u_1,p,z_0,w) \approx e(u_0,p,z_0,w)+\frac{\partial e(u_0,p,z_0,w)}{\partial u}(u_1-u_0)$$

$e(u_1, p, tz_0, w) \approx e(u_0, p, tz_0, w) + \frac{\partial e(u_0,p,tz_0,w)}{\partial u}(u_1 - u_0)$

Substituting the above expansions into (A1) and canceling then implies:

$$\frac{\partial e(u_0, p, z_0, w)}{\partial u} > \frac{\partial e(u_0, p, tz_0, w)}{\partial u}. \qquad \text{(A2)}$$

The relationship in (A2) is a direct consequence of EV $>$ CV. This derivative of the expenditure function with respect to utility is the incremental expenditure required to raise utility by one unit at a given level of the public good. Therefore, the inequality in (A2) means that it costs more money to achieve a one unit increase in utility when the public good is at the lower level $z_0$ than when it is at the higher level $tz_0$ (when $t > 1$). The economic intuition is as follows. Utility levels increase due to the increase in private as well as public goods. However, since public goods are exogeneous, they cannot be directly increased through an increase in the expenditure. When public good provision is low, individuals spend additional income on private substitutes to maintain the same level of utility ($u_0$). This additional expenditure makes up for the low level of the public good and means that improving utility by one unit also requires additional marginal spending relative to a higher level of public good provision. In other words, when the public good provision is higher, less spending on substitute private goods is needed, so raising utility costs less.

Next, we show that the private and the public goods that satisfy the independent homogeneity assumption are (imperfect) substitutes. Recall that:

$$t^{\theta} u(\boldsymbol{x}; \boldsymbol{z}, \boldsymbol{w}) = u(\boldsymbol{x}; t\boldsymbol{z}, \boldsymbol{w}) \qquad \text{(A3)}$$

$$t^{\eta} u(\boldsymbol{x}; \boldsymbol{z}, \boldsymbol{w}) = u(t\boldsymbol{x}; \boldsymbol{z}, \boldsymbol{w}). \qquad \text{(A4)}$$

Then, starting with (A3): $t^{\theta}u(\boldsymbol{x}; \boldsymbol{z}, \boldsymbol{w}) = u(\boldsymbol{x}; t\boldsymbol{z}, \boldsymbol{w}) \Leftrightarrow t^{(\theta/\eta)\eta}u(\boldsymbol{x}; \boldsymbol{z}, \boldsymbol{w}) = u(\boldsymbol{x}; t\boldsymbol{z}, \boldsymbol{w})$. Next, using (A4), we get $u\left(t^{\theta/\eta}\boldsymbol{x}; \boldsymbol{z}, \boldsymbol{w}\right) = u(\boldsymbol{x}; t\boldsymbol{z}, \boldsymbol{w})$. This last relationship shows that an increase in public goods (that satisfies the independent homogeneity condition) can be substituted by an increase in the private good and implies that these are (imperfect) substitutes.

**Proof: Independent homogeneity excludes perfect substitutes.**

For two goods to be perfect substitutes, their marginal rate of substitution (MRS) must be constant. Consider the special case of an independently homogenous utility function with a single public good $z$ such that $t^{\eta}u(\boldsymbol{x}; z) = u(t\boldsymbol{x}; z)$ and $t^{\theta}u(\boldsymbol{x}; z) = u(\boldsymbol{x}; tz)$. Also, let $\tilde{\boldsymbol{x}} = t\boldsymbol{x}$ and $\tilde{z} = tz$ (with $t \neq 1$). Then, differentiating finds with respect to some private good $x_i$:

$$t^{\eta}\frac{\partial u}{\partial x_i} = \frac{\partial u}{\partial \tilde{x}_i}\frac{\partial \tilde{x}_i}{\partial x_i} \Leftrightarrow t^{\eta}\frac{\partial u}{\partial x_i} = \frac{\partial u}{\partial \tilde{x}_i}t \Leftrightarrow t^{\eta-1}\frac{\partial u}{\partial x_i} = \frac{\partial u}{\partial \tilde{x}_i} \Leftrightarrow t^{\eta-1}\frac{\frac{\partial u}{\partial x_i}}{\frac{\partial u}{\partial z}} = \frac{\frac{\partial u}{\partial \tilde{x}_i}}{\frac{\partial u}{\partial z}} \Leftrightarrow \frac{\frac{\partial u}{\partial x_i}}{\frac{\partial u}{\partial z}} = t^{1-\eta}\frac{\frac{\partial u}{\partial \tilde{x}_i}}{\frac{\partial u}{\partial z}}$$

and, similarly, for $z$,

$$t^{\theta}\frac{\partial u}{\partial z} = \frac{\partial u}{\partial \tilde{z}}\frac{\partial \tilde{z}}{\partial z} \Leftrightarrow t^{\theta}\frac{\partial u}{\partial z} = \frac{\partial u}{\partial \tilde{z}}t \Leftrightarrow t^{\theta-1}\frac{\partial u}{\partial z} = \frac{\partial u}{\partial \tilde{z}} \Leftrightarrow t^{1-\theta}\frac{\frac{\partial u}{\partial x_i}}{\frac{\partial u}{\partial z}} = \frac{\frac{\partial u}{\partial x_i}}{\frac{\partial u}{\partial \tilde{z}}} \Leftrightarrow \frac{\frac{\partial u}{\partial x_i}}{\frac{\partial u}{\partial z}} = t^{\theta-1}\frac{\frac{\partial u}{\partial x_i}}{\frac{\partial u}{\partial \tilde{z}}}.$$

Then, the MRS at $(x_i, z)$ equals $\frac{\partial u}{\partial x_i}\Big/\frac{\partial u}{\partial z}$, and setting equal from both expressions above finds:

$$\frac{\frac{\partial u}{\partial x_i}}{\frac{\partial u}{\partial z}} = t^{1-\eta}\frac{\frac{\partial u}{\partial \tilde{x}_i}}{\frac{\partial u}{\partial z}} = t^{\theta-1}\frac{\frac{\partial u}{\partial x_i}}{\frac{\partial u}{\partial \tilde{z}}} \Longrightarrow MRS_{(x_i,z)} = t^{1-\eta}MRS_{(\tilde{x}_i,z)} = t^{\theta-1}MRS_{(x_i,\tilde{z})} \quad \text{[A]}$$

and thus, the MRS at $(x_i, z)$ is different from the MRS at $(\tilde{x}_i, z)$ and $(\tilde{x}_i, \tilde{z})$ since there does not exist a real constant $c = MRS_{(x_i,z)} = MRS_{(\tilde{x}_i,z)} = MRS_{(x_i,\tilde{z})}$ that satisfies expression [A] when $\eta \neq 1$ or $\theta \neq 1$. Now, consider now the special case of $\eta = 1$ and $\theta = 1$, and it follows that:

$$u(t\boldsymbol{x}; t^{-1}z) = (t^{1}t^{-1})u(\boldsymbol{x}; z) = u(\boldsymbol{x}; z)$$

Next, let $\tilde{\boldsymbol{x}} = t\boldsymbol{x}$ and $\tilde{z} = t^{-1}z$, and differentiating by $x_i$ finds:

$$\frac{\partial u(t\boldsymbol{x};t^{-1}z)}{\partial \tilde{x}_i}\frac{\partial \tilde{x}_i}{\partial x_i}=\frac{\partial u(\boldsymbol{x};z)}{\partial x_i}\Rightarrow t\frac{\partial u(\boldsymbol{x};t^{-1}z)}{\partial \tilde{x}_i}=\frac{\partial u(\boldsymbol{x};z)}{\partial x_i}.$$

Similarly, differentiating by $z$ finds:

$$\frac{\partial u(t\boldsymbol{x};t^{-1}z)}{\partial \tilde{z}}\frac{\partial \tilde{z}_1}{\partial z}=\frac{\partial u(\boldsymbol{x};z)}{\partial z}\Rightarrow t^{-1}\frac{\partial u(t\boldsymbol{x};t^{-1}z)}{\partial \tilde{z}}=\frac{\partial u(\boldsymbol{x};z)}{\partial z}.$$

Taking the ratios, we get $t^2 MRS_{(\tilde{x}_i,\tilde{z})} = MRS_{(x_i,z)}$. *Q.E.D.*

**Proof of Corollary 1.** A binding budget constraint can be rewritten as $m=\sum_{n=1}^{N} p_n x_n$. At the initial public good provision $m=\sum_{n=1}^{N} p_n x_n^*$ at $u_0$ and, similarly, $m=\sum_{n=1}^{N} p_n x_n^{**}$ at $u_1$. By Theorem 1, $t^{\phi} m=\sum_{n=1}^{N} p_n x_n'$ and $t^{-\phi} m=\sum_{n=1}^{N} p_n x_n''$ (with $\phi\equiv-\theta/\eta$). For compensating variation, it follows that: $CV=\left(1-t^{\phi}\right)m=m-t^{\phi}m=\sum_{n=1}^{N} p_n x_n^*-\sum_{n=1}^{N} p_n x_n'=\sum_{n=1}^{N}(p_n x_n^*-p_n x_n')=\sum_{n=1}^{N} p_n(x_n^*-x_n')=\sum_{n=1}^{N}\mathrm{CV}_n$.

For equivalent variation, it follows that: $EV=\left(t^{-\phi}-1\right)m=t^{-\phi}m-m=\sum_{n=1}^{N} p_n x_n''-\sum_{n=1}^{N} p_n x_n^{**}=\sum_{n=1}^{N}(p_n x_n''-p_n x_n^{**})=\sum_{n=1}^{N} p_n(x_n''-x_n^{**})=\sum_{n=1}^{N}\mathrm{EV}_n$. *Q.E.D.*

**Proof of Proposition 1.** Begin by assuming separability of the utility function and then prove the property of the expenditure function. To start,

$e(u,\boldsymbol{p};\boldsymbol{z})=\min_{\{x\}}\{\boldsymbol{p}\cdot\boldsymbol{x} : u(\boldsymbol{x};\boldsymbol{z})\geq u\}$ [Definition]

$\Leftrightarrow e(u,\boldsymbol{p};\boldsymbol{z})=\min_{\{x\}}\{\boldsymbol{p}\cdot\boldsymbol{x} : u_1(\boldsymbol{x})+u_2(\boldsymbol{z})\geq u\}$ [Separability assumption]

$\Leftrightarrow e(u,\boldsymbol{p};\boldsymbol{z})=\min_{\{x\}}\{\boldsymbol{p}\cdot\boldsymbol{x} : u_1(\boldsymbol{x})\geq u-u_2(\boldsymbol{z})\}$

$\Leftrightarrow e(u,\boldsymbol{p};\boldsymbol{z})=e(u-u_2(\boldsymbol{z}),\boldsymbol{p})$ [Definition].

Then, assume that the expenditure function can be expressed as $e(u,\boldsymbol{p};\boldsymbol{z})=e(u-u_2(\boldsymbol{z}),\boldsymbol{p})$ and show that the utility function is additively separable in public goods. To start,

$e(u-u_2(\boldsymbol{z}),\boldsymbol{p})=\min_{\{x\}}\{\boldsymbol{p}\cdot\boldsymbol{x} : u_1(\boldsymbol{x})\geq u-u_2(\boldsymbol{z})\}$ [Definition]

$\Leftrightarrow e(u - u_2(\boldsymbol{z}), \boldsymbol{p}) = \min_{\{x\}} \{\boldsymbol{p} \cdot \boldsymbol{x} : u_1(\boldsymbol{x}) + u_2(\boldsymbol{z}) \geq u \}$

$\Leftrightarrow e(u - u_2(\boldsymbol{z}), \boldsymbol{p}) = \min_{\{x\}} \{\boldsymbol{p} \cdot \boldsymbol{x} : u(\boldsymbol{x}; \boldsymbol{z}) \geq u \}$ [Separability definition]. *Q.E.D.*

**Proof of Proposition 2.** By Espinosa and Prada (2012)'s Corollary 1 and 2, respectively, $v_1(m, \boldsymbol{p}) = m^{\gamma} e(1, \boldsymbol{p})^{-\gamma}$ and $v_1(m, \boldsymbol{p}) = m^{\gamma} v_1(1, \boldsymbol{p})$, when $u_1(\boldsymbol{x})$ is homogeneous of degree $\gamma \neq 0$. Then, it follows $v_1(1, \boldsymbol{p}) = e(1, \boldsymbol{p})^{-\gamma}$ and $v_1(1, \boldsymbol{p})^{-1/\gamma} = e(1, \boldsymbol{p})$. Next,

$e(u, \boldsymbol{p}; \boldsymbol{z}) = e(u - u_2(\boldsymbol{z}), \boldsymbol{p})$ [Proposition 1]

$\Leftrightarrow e(u, \boldsymbol{p}; \boldsymbol{z}) = e(\tilde{u}, \boldsymbol{p})$ [Definition $\tilde{u} = u - u_2(\boldsymbol{z})$]

$\Leftrightarrow e(u, \boldsymbol{p}; \boldsymbol{z}) = \tilde{u}^{1/\gamma} e(1, \boldsymbol{p})$ [Espinosa and Prada (2012)'s Corollary 1]

$\Leftrightarrow e(u, \boldsymbol{p}; \boldsymbol{z}) = \tilde{u}^{1/\gamma} v_1(1, \boldsymbol{p})^{-1/\gamma} \Leftrightarrow e(u, \boldsymbol{p}; \boldsymbol{z}) = \left[\frac{u - u_2(\boldsymbol{z})}{v_1(\mathbf{1}, \mathbf{p})}\right]^{1/\gamma}$ [Definition]. *Q.E.D.*

# Appendix B

We assume a homogeneous utility function by letting $g(\cdot)$ be the identity function and provide results related to indirect utility function and the expenditure function. We also derive the certainty equivalence of lotteries over public goods.

To start, Proposition B1 shows the relationship between direct and indirect utility under continuity in our public goods model. Proposition B1 is the only result in this study that does not utilize a homogeneity assumption and thus applies to both jointly and independently homogeneous functions.

**Proposition B1.** If the utility function is strictly increasing, continuous, and quasiconcave, then $u(\boldsymbol{x}; \boldsymbol{z}, \boldsymbol{w}) = \min_{\boldsymbol{p}} \{v(m, \boldsymbol{p}; \boldsymbol{z}, \boldsymbol{w}) : \boldsymbol{p} \cdot \boldsymbol{x} = m\}$.

**Proof of Proposition B1.** All proofs for Appendix B appear at the end of this appendix.

Example B1 below provides closed-form Marshallian demands and an indirect utility function for a homogeneous utility function. For our examples, let $\boldsymbol{z} = (z_1\ z_2)$ where the first subscript indicates that these public goods are subject to the homogeneity assumption and let the $\boldsymbol{w}$ vector be empty. Example B1 is not weakly separable (and thus not strongly separable) because the MRS between the two private goods $(x_1, x_2)$ is given by $\text{MRS}_{x_1,x_2} = \frac{z_1}{z_2}\left(\frac{x_1}{x_2}\right)^{\alpha-1}$. Example B1 illustrates the "if" statements of Propositions B2a and B2b that follow, since the utility function in the example is both jointly and independently homogeneous, while the "only if" parts are left for the reader. Specifically, Propositions B2a and B2b relate the degree of homogeneity of a jointly and independently homogeneous utility function, respectively, to the indirect utility function. These if-and-only-if-style propositions demonstrate how maximized utility scales with respect to exogenous income and the provision of public goods.

Then, Propositions B2c and B2d separately apply each of the necessary condition clauses of independent homogeneity and doing so leads to Corollary B1 regarding Marshallian demand with respect to public goods under homogeneity. Specifically, Corollary B1 shows that if a strictly quasiconcave utility function is homogeneous in only the public goods $(\boldsymbol{z})$ then Marshallian demand is homogeneous of degree zero in the public goods, and Example B1 also demonstrates this property. The utility scaling in Propositions B2a through B2d requires the cardinal properties of a homogeneous utility function.

**Example B1:** Let $u(\boldsymbol{x}; \boldsymbol{z}) = x_1^{\alpha} z_1 + x_2^{\alpha} z_2$ and thus $t^{\alpha+1} u(\boldsymbol{x}; \boldsymbol{z}) = u(t\boldsymbol{x}; t\boldsymbol{z})$; that is, $u(\boldsymbol{x}; \boldsymbol{z})$ is jointly homogeneous of degree $\alpha + 1$. This utility function is also independently homogeneous of degrees 1 and $\alpha$ in the vectors $\boldsymbol{x}$ and $\boldsymbol{z}$, respectively. The first-order conditions of the UMP lead to $\frac{x_1}{x_2} = \left[\frac{P_1 z_2}{P_2 z_1}\right]^{1/(\alpha-1)} \equiv k(\boldsymbol{p}; \boldsymbol{z}) = k$ and restrict $\alpha \neq 0, 1$. Note that $k(\boldsymbol{p}; t\boldsymbol{z}) = k(\boldsymbol{p}; \boldsymbol{z})$. Next, the Marshallian demands are $x_1(m, \boldsymbol{p}; \boldsymbol{z}) = \frac{mk}{P_1 k + P_2}$ and $x_2(m, \boldsymbol{p}; \boldsymbol{z}) = \frac{m}{P_1 k + P_2}$, and have the properties $t x_n(m, \boldsymbol{p}; \boldsymbol{z}) = x_n(tm, \boldsymbol{p}; t\boldsymbol{z})$ as well as $x_n(m, \boldsymbol{p}; \boldsymbol{z}) = x_n(m, \boldsymbol{p}; t\boldsymbol{z})$ for $n = 1,2$. Thus, the indirect utility function has the closed-form solution $v(m, \boldsymbol{p}; \boldsymbol{z}) = (k^\alpha z_1 + z_2)\left(\frac{m}{P_1 k + P_2}\right)^\alpha$ and has the property that $t^{\alpha+1} v(m, \boldsymbol{p}; \boldsymbol{z}) = v(tm, \boldsymbol{p}; t\boldsymbol{z})$.

**Proposition B2a.** The utility function $u(\boldsymbol{x}; \boldsymbol{z}, \boldsymbol{w})$ is jointly homogeneous of degree $\gamma$ (see definition in Section 2) if and only if the indirect utility function $v(m, \boldsymbol{p}; \boldsymbol{z}, \boldsymbol{w})$ is homogeneous of degree $\gamma$ in income $(m)$ and public goods $(\boldsymbol{z}_1)$; that is, $t^\gamma v(m, \boldsymbol{p}; \boldsymbol{z}, \boldsymbol{w}) = v(tm, \boldsymbol{p}; t\boldsymbol{z}, \boldsymbol{w})$ for all $t > 0$.

**Proposition B2b.** The utility function $u(\boldsymbol{x}; \boldsymbol{z}, \boldsymbol{w})$ is independently homogeneous (see definition in Section 2) if and only if the indirect utility function $v(m, \boldsymbol{p}; \boldsymbol{z}, \boldsymbol{w})$ is homogeneous of degree $\eta + \theta$ in income and public goods $(\boldsymbol{z})$; that is, $t^{\eta+\theta} v(m, \boldsymbol{p}; \boldsymbol{z}, \boldsymbol{w}) = v(tm, \boldsymbol{p}; t\boldsymbol{z}, \boldsymbol{w})$ for all $t > 0$.

**Proposition B2c.** The utility function $u(\boldsymbol{x}; \boldsymbol{z}, \boldsymbol{w})$ is homogeneous in the private goods such that $t^\eta u(\boldsymbol{x}, \boldsymbol{z}, \boldsymbol{w}) = u(t\boldsymbol{x}, \boldsymbol{z}, \boldsymbol{w})$ for all $t > 0$ if and only if $t^\eta v(m, \boldsymbol{p}; \boldsymbol{z}, \boldsymbol{w}) = v(tm, \boldsymbol{p}; \boldsymbol{z}, \boldsymbol{w})$.

**Proposition B2d.** The utility function $u(\boldsymbol{x}; \boldsymbol{z}, \boldsymbol{w})$ is homogeneous in the public goods $(\boldsymbol{z})$ such that $t^\theta u(\boldsymbol{x}, \boldsymbol{z}, \boldsymbol{w}) = u(\boldsymbol{x}, t\boldsymbol{z}, \boldsymbol{w})$ for all $t > 0$ if and only if $t^\theta v(m, \boldsymbol{p}; \boldsymbol{z}, \boldsymbol{w}) = v(m, \boldsymbol{p}; t\boldsymbol{z}, \boldsymbol{w})$.

**Corollary B1.** If the utility function is homogeneous in the public goods ($\boldsymbol{z}$), such that $t^{\theta}u(\boldsymbol{x},\boldsymbol{z},\boldsymbol{w}) = u(\boldsymbol{x},t\boldsymbol{z},\boldsymbol{w})$ for all $t > 0$ and $u(\cdot)$ is strictly quasiconcave, then Marshallian demands are homogeneous of degree zero in public goods where $x_n(m,\boldsymbol{p};\boldsymbol{z},\boldsymbol{w}) = x_n(m,\boldsymbol{p};t\boldsymbol{z},\boldsymbol{w})$ for all $t > 0$ for all $n = 1,\ldots,N$.

Next, we explore properties of the expenditure function given a jointly homogeneous utility function. With an if-and-only-if structure proof, Proposition B3 formalizes Example B1's finding that relates the utility function's joint degree of homogeneity to scaling utility and the public goods. Corollary B2 follows directly from Proposition B3 by setting $\gamma = 1$. Corollary B2 is easy to interpret: the utility function is jointly homogeneous of degree one, equivalent to assuming homothetic preferences (Varian, 2002), if and only if the expenditure function is jointly homogeneous of degree one in utility and public goods. Corollary B2 also proves that Hicksian demand is homogeneous of degree zero in utility and public goods when the utility function is strictly quasiconcave. The CES utility function is an important example of a utility function with homogeneity of degree 1 while noting that CES functions are additively separable in a monotonic transformation and thus are strongly separable (Deaton and Muellbauer, 1980). In some contexts, the degree of homogeneity of utility function dictates whether an agent is a risk-taker, risk-neutral, or risk-averse. Homogeneous utility is also employed in asset pricing models with atemporal, non-expected utility theories (Epstein and Zin, 1989).

**Example B1 (concluded):** The utility function $u(\boldsymbol{x};\boldsymbol{z}) = x_1^{\alpha}z_1 + x_2^{\alpha}z_2$ is jointly homogeneous of degree $\alpha + 1$ and the expenditure function is $e(u,\boldsymbol{p};\boldsymbol{z}) = (kp_1 + p_2)\left[\frac{u}{k^{\alpha}z_1+z_2}\right]^{1/\alpha}$ with the property $te(u,\boldsymbol{p};\boldsymbol{z}) = e(t^{\alpha+1}u,\boldsymbol{p};t\boldsymbol{z})$.

**Proposition B3.** The utility function $u(\boldsymbol{x}; \boldsymbol{z}, \boldsymbol{w})$ is jointly homogeneous of degree $\gamma$ if and only if the expenditure function $e(u, \boldsymbol{p}; \boldsymbol{z}, \boldsymbol{w})$ satisfies the following property: $te(u, \boldsymbol{p}; \boldsymbol{z}, \boldsymbol{w}) = e(t^{\gamma}u, \boldsymbol{p}; t\boldsymbol{z}, \boldsymbol{w})$ for all $t > 0$.

**Corollary B2.** The utility function $u(\boldsymbol{x}; \boldsymbol{z}, \boldsymbol{w})$ is jointly homogeneous of degree 1 in $\boldsymbol{x}$ and $\boldsymbol{z}$ if and only if the expenditure function $e(u, \boldsymbol{p}; \boldsymbol{z}, \boldsymbol{w})$ is homogeneous of degree 1 in utility and public goods; that is, $te(u, \boldsymbol{p}; \boldsymbol{z}, \boldsymbol{w}) = e(tu, \boldsymbol{p}; t\boldsymbol{z}, \boldsymbol{w})$ for all $t > 0$. Furthermore, if the utility function is jointly homogeneous of degree 1 and $u(\cdot)$ is strictly quasiconcave, then the Hicksian demand is homogeneous of degree 0 such that $x_n^h(u, \boldsymbol{p}; \boldsymbol{z}, \boldsymbol{w}) = x_n^h(tu, \boldsymbol{p}; t\boldsymbol{z}, \boldsymbol{w})$ for all $t > 0$ for all $n = 1, \dots, N$.

We conclude the analysis of homogeneous utility by deriving the certainty equivalence ($CE$) of lotteries over public goods and thus provide a measure of the curvature of $u(\cdot)$ related to the preferences parameters $\theta$ and $\eta$.

To begin, we express the certainty equivalence as $CE = (1 - r)m$, where $r > 0$ is the risk premium expressed as a percentage of income. Let $\tilde{\boldsymbol{z}}$ be the amount of the public goods that the consumer has with certainty. The lottery occurs over two possible outcomes $\tilde{\boldsymbol{z}}_L$ and $\tilde{\boldsymbol{z}}_H$. Let the "low" outcome be given by $\tilde{\boldsymbol{z}}_L \equiv \delta_L \tilde{\boldsymbol{z}}$ with $\delta_L \in [0,1)$ and the "high" outcome be given by $\tilde{\boldsymbol{z}}_H \equiv \delta_H \tilde{\boldsymbol{z}}$ with $\delta_H \in (1, \infty)$ such that $\tilde{\boldsymbol{z}}_L < \tilde{\boldsymbol{z}} < \tilde{\boldsymbol{z}}_H$. Also, let the probabilities assigned to $\tilde{\boldsymbol{z}}_L$ and $\tilde{\boldsymbol{z}}_H$ be denoted $w$ and $1 - w$, respectively, with $w \in [0,1]$.

The goal is then to find $r > 0$ associated with $\tilde{\boldsymbol{z}}$ and thus the $CE$ expression written in terms of the indirect utility function is as follows:

$$v\big((1-r)m,\boldsymbol{p};\tilde{\boldsymbol{z}},\boldsymbol{w}\big)=wv(m,\boldsymbol{p};\tilde{\boldsymbol{z}}_L,\boldsymbol{w})+(1-w)v(m,\boldsymbol{p};\tilde{\boldsymbol{z}}_H,\boldsymbol{w}) \tag{B1}$$

and we proceed by applying the definitions and then Propositions B2c and B2d to the left-hand and right-hand sides, respectively, of equation (B1) under the assumption of independent homogeneity. After cancelling $v(m,\boldsymbol{p};\tilde{\boldsymbol{z}},\boldsymbol{w})$ from both sides and rearranging, we find an expression for $r$ that is then substituted into the certainty equivalence definition to yield:

$$CE=\left[w\delta_L^{\theta}+(1-w)\delta_H^{\theta}\right]^{1/\eta}m \tag{B2}$$

where $CE$ is positive.

Regarding the comparative statics in equation (B2) assuming $\theta>0$ and $\eta>0$, we find that $CE$ is increasing in income ($m$), the probability of the low outcome ($w$), the value of the low outcome ($\delta_L$), the value of the high outcome ($\delta_H$), and the public good preference parameter ($\theta$). The $CE$ is increasing in $\theta$ since a high value for public goods means the consumer place a premium on securing their provision of public goods. Meanwhile, $CE$ is decreasing in private good preference parameter ($\eta$) and this occurs because the more a consumer values the utility from private goods then the less the consumer is willing to pay to avoid the lottery for the public goods. That is, we find opposite effects of $\theta$ and $\eta$ on the certainty equivalence and this finding foreshadows results in subsequent sections under homothetic utility. However, the $CE$ expression in equation (B2) does not extend to homothetic utility since Propositions B2c and B2d only hold for homogeneous utility functions.

**Proof of Proposition B1.** Fix $\boldsymbol{z}$ and $\boldsymbol{w}$, then let $\tilde{u}(\boldsymbol{x})=u(\boldsymbol{x};\boldsymbol{z},\boldsymbol{w})$. Then, the function $\tilde{u}(\cdot)$ is strictly increasing in $\boldsymbol{x}$ and it follows that the upper contour level sets of $\tilde{u}$ are convex (since $u(\cdot)$ is continuous and quasiconcave function). Then, $\tilde{u}(\cdot)$ satisfies conditions C.2 and C.3 from Espinosa and Prada (2012) and thus,

$\tilde{u}(\boldsymbol{x}) = \min_p\{\tilde{v}(m, \boldsymbol{p}) : \boldsymbol{p} \cdot \boldsymbol{x} = m\}$ [Definition, Proposition 2: Espinosa and Prada (2012)] where $\tilde{v}(m, \boldsymbol{p}) = \max_{\boldsymbol{x}}\{\tilde{u}(\boldsymbol{x}) : \boldsymbol{p} \cdot \boldsymbol{x} = m\}$ [Definition]

$\Leftrightarrow \tilde{v}(m, \boldsymbol{p}) = \max_{\boldsymbol{x}}\{u(\boldsymbol{x}; \boldsymbol{z}, \boldsymbol{w}) : \boldsymbol{p} \cdot \boldsymbol{x} = m\}$ [Since $\tilde{u}(\boldsymbol{x}) = u(\boldsymbol{x}; \boldsymbol{z})$]

$\Leftrightarrow \tilde{v}(m, \boldsymbol{p}) = v(m, \boldsymbol{p}; \boldsymbol{z}, \boldsymbol{w})$ [Definition]

$\Leftrightarrow u(\boldsymbol{x}; \boldsymbol{z}, \boldsymbol{w}) = \min_p\{v(m, \boldsymbol{p}; \boldsymbol{z}, \boldsymbol{w}) : \boldsymbol{p} \cdot \boldsymbol{x} = m\}$ [Definition]. *Q.E.D.*

**Proof of Proposition B2a.** Begin by assuming that the utility function is homogeneous of degree $\gamma$ and show that the indirect utility function is homogeneous of degree $\gamma$ in income and public goods. To start, $v(m, \boldsymbol{p}; \boldsymbol{z}, \boldsymbol{w}) = \max_{\boldsymbol{x}}\{u(\boldsymbol{x}; \boldsymbol{z}, \boldsymbol{w}) : \boldsymbol{p} \cdot \boldsymbol{x} \le m\}$ [Definition]

$\Leftrightarrow t^\gamma v(m, \boldsymbol{p}; \boldsymbol{z}, \boldsymbol{w}) = t^\gamma \max_{\boldsymbol{x}}\{u(\boldsymbol{x}; \boldsymbol{z}, \boldsymbol{w}) : \boldsymbol{p} \cdot \boldsymbol{x} \le m\}$

$\Leftrightarrow t^\gamma v(m, \boldsymbol{p}; \boldsymbol{z}, \boldsymbol{w}) = \max_{\boldsymbol{x}}\{t^\gamma u(\boldsymbol{x}; \boldsymbol{z}, \boldsymbol{w}) : \boldsymbol{p} \cdot \boldsymbol{x} \le m\}$

$\Leftrightarrow t^\gamma v(m, \boldsymbol{p}; \boldsymbol{z}, \boldsymbol{w}) = \max_{\boldsymbol{x}}\{u(t\boldsymbol{x}; t\boldsymbol{z}, \boldsymbol{w}) : \boldsymbol{p} \cdot \boldsymbol{x} \le m\}$ [Assumption]

$\Leftrightarrow t^\gamma v(m, \boldsymbol{p}; \boldsymbol{z}, \boldsymbol{w}) = \max_{\tilde{\boldsymbol{x}}}\left\{u(\tilde{\boldsymbol{x}}; t\boldsymbol{z}, \boldsymbol{w}) : \boldsymbol{p} \cdot \left(\frac{1}{t}\tilde{\boldsymbol{x}}\right) \le m\right\}$ [Unit conversion : $\tilde{\boldsymbol{x}} = t\boldsymbol{x}$]

$\Leftrightarrow t^\gamma v(m, \boldsymbol{p}; \boldsymbol{z}, \boldsymbol{w}) = \max_{\tilde{\boldsymbol{x}}}\{u(\tilde{\boldsymbol{x}}; t\boldsymbol{z}, \boldsymbol{w}) : \boldsymbol{p} \cdot \tilde{\boldsymbol{x}} \le tm\}$

$\Leftrightarrow t^\gamma v(m, \boldsymbol{p}; \boldsymbol{z}, \boldsymbol{w}) = v(tm, \boldsymbol{p}; t\boldsymbol{z}, \boldsymbol{w})$ [Definition].

Next, assume that the indirect utility function is homogeneous of degree $\gamma$ in income and public goods and show that the utility function is homogeneous of degree $\gamma$. Start with the result from Proposition B1 and proceed as follows:

$u(\boldsymbol{x}; \boldsymbol{z}, \boldsymbol{w}) = \min_{\boldsymbol{p}}\{v(m, \boldsymbol{p}; \boldsymbol{z}, \boldsymbol{w}) : \boldsymbol{p} \cdot \boldsymbol{x} = m\}$ [Proposition B1]

$\Leftrightarrow t^\gamma u(\boldsymbol{x}; \boldsymbol{z}, \boldsymbol{w}) = t^\gamma \min_{\boldsymbol{p}}\{v(m, \boldsymbol{p}; \boldsymbol{z}, \boldsymbol{w}) : \boldsymbol{p} \cdot \boldsymbol{x} = m\}$

$\Leftrightarrow t^\gamma u(\boldsymbol{x}; \boldsymbol{z}, \boldsymbol{w}) = \min_{\boldsymbol{p}}\{t^\gamma v(m, \boldsymbol{p}; \boldsymbol{z}, \boldsymbol{w}) : \boldsymbol{p} \cdot \boldsymbol{x} = m\}$

$\Leftrightarrow t^{\gamma}u(\boldsymbol{x};\boldsymbol{z},\boldsymbol{w}) = \min_{\boldsymbol{p}}\{v(tm,\boldsymbol{p};t\boldsymbol{z},\boldsymbol{w}):\boldsymbol{p}\cdot\boldsymbol{x} = m\}$ [Assumption]

$\Leftrightarrow t^{\gamma}u(\boldsymbol{x};\boldsymbol{z},\boldsymbol{w}) = \min_{\boldsymbol{p}}\left\{t^{0}v\left(m,\frac{1}{t}\boldsymbol{p};t\boldsymbol{z},\boldsymbol{w}\right):\boldsymbol{p}\cdot\boldsymbol{x} = m\right\}$ [Indirect utility homogeneous of degree zero in prices and income]

$\Leftrightarrow t^{\gamma}u(\boldsymbol{x};\boldsymbol{z},\boldsymbol{w}) = \min_{\widetilde{\boldsymbol{p}}}\{v(m,\widetilde{\boldsymbol{p}};t\boldsymbol{z},\boldsymbol{w}):t\widetilde{\boldsymbol{p}}\cdot\boldsymbol{x} = m\}$ [Unit conversion such that $\widetilde{\boldsymbol{p}} = \frac{1}{t}\boldsymbol{p}$]

$\Leftrightarrow t^{\gamma}u(\boldsymbol{x};\boldsymbol{z},\boldsymbol{w}) = \min_{\widetilde{\boldsymbol{p}}}\{v(m,\widetilde{\boldsymbol{p}};t\boldsymbol{z},\boldsymbol{w}):\widetilde{\boldsymbol{p}}\cdot t\boldsymbol{x} = m\}$

$\Leftrightarrow t^{\gamma}u(\boldsymbol{x};\boldsymbol{z},\boldsymbol{w}) = u(t\boldsymbol{x};t\boldsymbol{z},\boldsymbol{w})$ [Proposition A1]. *Q.E.D.*

**Proof of Proposition B2b.** Begin by assuming that the utility function is independently homogeneous and show that the indirect utility function is homogeneous of degree $\theta$ in income and public goods. To start, $v(m,\boldsymbol{p};\boldsymbol{z},\boldsymbol{w}) = \max_{\boldsymbol{x}}\{u(\boldsymbol{x};\boldsymbol{z},\boldsymbol{w}):\boldsymbol{p}\cdot\boldsymbol{x} \leq m\}$ [Definition]

$\Leftrightarrow t^{\eta+\theta}v(m,\boldsymbol{p};\boldsymbol{z},\boldsymbol{w}) = t^{\eta+\theta}\max_{\boldsymbol{x}}\{u(\boldsymbol{x};\boldsymbol{z},\boldsymbol{w}):\boldsymbol{p}\cdot\boldsymbol{x} \leq m\}$

$\Leftrightarrow t^{\eta+\theta}v(m,\boldsymbol{p};\boldsymbol{z},\boldsymbol{w}) = t^{\eta}t^{\theta}\max_{\boldsymbol{x}}\{u(\boldsymbol{x};\boldsymbol{z},\boldsymbol{w}):\boldsymbol{p}\cdot\boldsymbol{x} \leq m\}$

$\Leftrightarrow t^{\eta+\theta}v(m,\boldsymbol{p};\boldsymbol{z},\boldsymbol{w}) = \max_{\boldsymbol{x}}\{t^{\eta}t^{\theta}u(\boldsymbol{x};\boldsymbol{z},\boldsymbol{w}):\boldsymbol{p}\cdot\boldsymbol{x} \leq m\}$

$\Leftrightarrow t^{\eta+\theta}v(m,\boldsymbol{p};\boldsymbol{z},\boldsymbol{w}) = \max_{\boldsymbol{x}}\{u(t\boldsymbol{x};t\boldsymbol{z},\boldsymbol{w}):\boldsymbol{p}\cdot\boldsymbol{x} \leq m\}$ [Assumption]

$\Leftrightarrow t^{\eta+\theta}v(m,\boldsymbol{p};\boldsymbol{z},\boldsymbol{w}) = \max_{\widetilde{\boldsymbol{x}}}\left\{u(\widetilde{\boldsymbol{x}};t\boldsymbol{z},\boldsymbol{w}):\boldsymbol{p}\cdot\left(\frac{1}{t}\widetilde{\boldsymbol{x}}\right) \leq m\right\}$ [Unit conversion : $\widetilde{\boldsymbol{x}} = t\boldsymbol{x}$]

$\Leftrightarrow t^{\eta+\theta}v(m,\boldsymbol{p};\boldsymbol{z},\boldsymbol{w}) = \max_{\widetilde{\boldsymbol{x}}}\{u(\widetilde{\boldsymbol{x}};t\boldsymbol{z},\boldsymbol{w}):\boldsymbol{p}\cdot\widetilde{\boldsymbol{x}} \leq tm\}$

$\Leftrightarrow t^{\eta+\theta}v(m,\boldsymbol{p};\boldsymbol{z},\boldsymbol{w}) = v(tm,\boldsymbol{p};t\boldsymbol{z},\boldsymbol{w})$ [Definition].

Next, assume that the indirect utility function is homogeneous of degree $\eta+\theta$ in income and public goods vector $\boldsymbol{z}$ and show that the utility function is homogeneous of degree $\eta+\theta$. Start with the result from Proposition B1 and proceed as follows:

$u(\boldsymbol{x};\boldsymbol{z},\boldsymbol{w}) = \min_{\boldsymbol{p}}\{v(m,\boldsymbol{p};\boldsymbol{z},\boldsymbol{w}):\boldsymbol{p}\cdot\boldsymbol{x} = m\}$ [Proposition B1]

$\Leftrightarrow t^{\eta+\theta}u(\boldsymbol{x};\boldsymbol{z},\boldsymbol{w}) = t^{\eta+\theta}\min_{\boldsymbol{p}}\{v(m,\boldsymbol{p};\boldsymbol{z},\boldsymbol{w}):\boldsymbol{p}\cdot\boldsymbol{x} = m\}$

$\Leftrightarrow t^{\eta+\theta}u(\boldsymbol{x};\boldsymbol{z},\boldsymbol{w}) = \min_{\boldsymbol{p}}\{t^{\eta+\theta}v(m,\boldsymbol{p};\boldsymbol{z},\boldsymbol{w}) : \boldsymbol{p}\cdot\boldsymbol{x} = m\}$

$\Leftrightarrow t^{\eta+\theta}u(\boldsymbol{x};\boldsymbol{z},\boldsymbol{w}) = \min_{\boldsymbol{p}}\{v(tm,\boldsymbol{p};t\boldsymbol{z},\boldsymbol{w}) : \boldsymbol{p}\cdot\boldsymbol{x} = m\}$ [Assumption]

$\Leftrightarrow t^{\eta+\theta}u(\boldsymbol{x};\boldsymbol{z},\boldsymbol{w}) = \min_{\boldsymbol{p}}\left\{t^{0}v\left(m,\frac{1}{t}\boldsymbol{p};t\boldsymbol{z},\boldsymbol{w}\right) : \boldsymbol{p}\cdot\boldsymbol{x} = m\right\}$ [Indirect utility homogeneous of degree zero in prices and income]

$\Leftrightarrow t^{\eta+\theta}u(\boldsymbol{x};\boldsymbol{z},\boldsymbol{w}) = \min_{\widetilde{\boldsymbol{p}}}\{v(m,\widetilde{\boldsymbol{p}};t\boldsymbol{z},\boldsymbol{w}) : t\widetilde{\boldsymbol{p}}\cdot\boldsymbol{x} = m\}$ [Unit conversion such that $\widetilde{\boldsymbol{p}} = \frac{1}{t}\boldsymbol{p}$]

$\Leftrightarrow t^{\eta+\theta}u(\boldsymbol{x};\boldsymbol{z},\boldsymbol{w}) = \min_{\widetilde{\boldsymbol{p}}}\{v(m,\widetilde{\boldsymbol{p}};t\boldsymbol{z},\boldsymbol{w}) : \widetilde{\boldsymbol{p}}\cdot t\boldsymbol{x} = m\}$

$\Leftrightarrow t^{\eta+\theta}u(\boldsymbol{x};\boldsymbol{z},\boldsymbol{w}) = u(t\boldsymbol{x};t\boldsymbol{z},\boldsymbol{w})$ [Proposition B1]. *Q.E.D.*

**Proof of Proposition B2c.** Follow Proof of Proposition B2b and simplify.

**Proof of Proposition B2d.** Follow Proof of Proposition B2b and simplify.

**Proof of Corollary B1.** Defined $\boldsymbol{x}^{*}$ to be the maximizer of $\max_{\boldsymbol{x}}\{u(\boldsymbol{x};\boldsymbol{z},\boldsymbol{w}) : \boldsymbol{p}\cdot\boldsymbol{x} \leq m\}$, and similarly, define $\boldsymbol{x}'$ to be the maximizer of $\max_{\boldsymbol{x}}\{u(\boldsymbol{x};t\boldsymbol{z},\boldsymbol{w}) : \boldsymbol{p}\cdot\boldsymbol{x} \leq m\}$. Then,

$v(m,\boldsymbol{p};\boldsymbol{z},\boldsymbol{w}) = \max_{\boldsymbol{x}}\{u(\boldsymbol{x};\boldsymbol{z},\boldsymbol{w}) : \boldsymbol{p}\cdot\boldsymbol{x} \leq m\}$ [Definition]

$\Leftrightarrow v(m,\boldsymbol{p};t\boldsymbol{z},\boldsymbol{w}) = \max_{\boldsymbol{x}}\{u(\boldsymbol{x};t\boldsymbol{z},\boldsymbol{w}) : \boldsymbol{p}\cdot\boldsymbol{x} \leq m\}$

$\Leftrightarrow v(m,\boldsymbol{p};t\boldsymbol{z},\boldsymbol{w}) = \max_{\boldsymbol{x}}\{t^{\theta}u(\boldsymbol{x};\boldsymbol{z},\boldsymbol{w}) : \boldsymbol{p}\cdot\boldsymbol{x} \leq m\}$ [Assumption]

$\Leftrightarrow v(m,\boldsymbol{p};t\boldsymbol{z},\boldsymbol{w}) = t^{\theta}\max_{\boldsymbol{x}}\{u(\boldsymbol{x};\boldsymbol{z},\boldsymbol{w}) : \boldsymbol{p}\cdot\boldsymbol{x} \leq m\}$ [Max operator property]

$\Leftrightarrow \max_{\boldsymbol{x}}\{u(\boldsymbol{x};t\boldsymbol{z},\boldsymbol{w}) : \boldsymbol{p}\cdot\boldsymbol{x} \leq m\} = t^{\theta}\max_{\boldsymbol{x}}\{u(\boldsymbol{x};\boldsymbol{z},\boldsymbol{w}) : \boldsymbol{p}\cdot\boldsymbol{x} \leq m\}$ [Definition]

and therefore $\boldsymbol{x}' = \boldsymbol{x}^{*}$ since scaling an objective function by $t^{\theta}$ does not affect the maximizer and the maximizers are unique by strict quasiconcavity. Alternatively, assuming a differentiable indirect utility function, this corollary can be proven by differentiating the indirect utility function and applying Roy's Identity. *Q.E.D.*

**Proof of Proposition B3.** Begin by assuming homogeneity of the utility function and then prove the scaling property of the expenditure function. Then,

$e(t^{\gamma}u, \boldsymbol{p}; t\boldsymbol{z}, \boldsymbol{w}) = \min_{\boldsymbol{x}}\{\boldsymbol{p} \cdot \boldsymbol{x}: u(\boldsymbol{x}; t\boldsymbol{z}, \boldsymbol{w}) \geq t^{\gamma}u\}$ [Definition]

$\Leftrightarrow e(t^{\gamma}u, \boldsymbol{p}; t\boldsymbol{z}, \boldsymbol{w}) = \min_{\boldsymbol{x}}\left\{\boldsymbol{p} \cdot \boldsymbol{x}: t^{\gamma}u\left(\frac{1}{t}\boldsymbol{x}; \boldsymbol{z}, \boldsymbol{w}\right) \geq t^{\gamma}u\right\}$ [Homogeneity of $u(\cdot)$]

$\Leftrightarrow e(t^{\gamma}u, \boldsymbol{p}; t\boldsymbol{z}, \boldsymbol{w}) = \min_{\boldsymbol{x}}\left\{\boldsymbol{p} \cdot \boldsymbol{x}: u\left(\frac{1}{t}\boldsymbol{x}; \boldsymbol{z}, \boldsymbol{w}\right) \geq u\right\}$ [Divide constraint by $t^{\gamma} > 0$]

$\Leftrightarrow e(t^{\gamma}u, \boldsymbol{p}; t\boldsymbol{z}, \boldsymbol{w}) = \min_{\widetilde{\boldsymbol{x}}}\{\boldsymbol{p} \cdot t\widetilde{\boldsymbol{x}}: u(\widetilde{\boldsymbol{x}}; \boldsymbol{z}, \boldsymbol{w}) \geq u\}$ [Unit conversion: $\widetilde{\boldsymbol{x}} = \frac{1}{t}\boldsymbol{x}$]

$\Leftrightarrow e(t^{\gamma}u, \boldsymbol{p}; t\boldsymbol{z}, \boldsymbol{w}) = t\min_{\widetilde{\boldsymbol{x}}}\{\boldsymbol{p} \cdot \widetilde{\boldsymbol{x}}: u(\widetilde{\boldsymbol{x}}; \boldsymbol{z}, \boldsymbol{w}) \geq u\}$

$\Leftrightarrow e(t^{\gamma}u, \boldsymbol{p}; t\boldsymbol{z}, \boldsymbol{w}) = te(u, \boldsymbol{p}; \boldsymbol{z}_1, \boldsymbol{z}_2)$ [Definition].

Next, assume the scaling property of the expenditure function and show the utility function property. Neary and Roberts (1980) show $u = v(e(u, \boldsymbol{p}; \boldsymbol{z}), \boldsymbol{p}; \boldsymbol{z})$, we omit $\boldsymbol{p}$ for notational simplicity; that is, $u = v(e(u; \boldsymbol{z}); \boldsymbol{z})$. In our model,

$v(m; \boldsymbol{z}, \boldsymbol{w}) = v(e(u; \boldsymbol{z}, \boldsymbol{w}); \boldsymbol{z}, \boldsymbol{w})$ [Identity: $m = e(u; \boldsymbol{z}, \boldsymbol{w})$]

$\Leftrightarrow v(tm; t\boldsymbol{z}, \boldsymbol{w}) = v(te(u; \boldsymbol{z}, \boldsymbol{w}); t\boldsymbol{z}, \boldsymbol{w})$

$\Leftrightarrow v(tm; t\boldsymbol{z}, \boldsymbol{w}) = v(e(t^{\gamma}u; t\boldsymbol{z}, \boldsymbol{w}); t\boldsymbol{z}, \boldsymbol{w})$ [By assumption $te(u; \boldsymbol{z}, \boldsymbol{w}) = e(t^{\gamma}u; t\boldsymbol{z}, \boldsymbol{w})$]

$\Leftrightarrow v(tm; t\boldsymbol{z}, \boldsymbol{w}) = t^{\gamma}u$ [Identity: $u = v(e(u; \boldsymbol{z}, \boldsymbol{w}); \boldsymbol{z}, \boldsymbol{w})$, so $v(e(t^{\gamma}u; t\boldsymbol{z}, \boldsymbol{w}); t\boldsymbol{z}, \boldsymbol{w}) = t^{\gamma}u$ by Proposition B2a.]

$\Leftrightarrow v(tm; t\boldsymbol{z}, \boldsymbol{w}) = t^{\gamma}v(m; \boldsymbol{z}, \boldsymbol{w})$ [Identity: $u = v(m; \boldsymbol{z}, \boldsymbol{w})$].

Thus, $v(\cdot)$ is homogeneous of degree $\gamma$ in income and public goods and therefore by Proposition B2a, $u(\cdot)$ is homogeneous of degree $\gamma$. *Q.E.D.*

**Proof of Corollary B2.** Start with Proposition B3 and set $\gamma = 1$. Define $\boldsymbol{x}^*$ to be the minimizer of $\min_{\boldsymbol{x}}\{\boldsymbol{p} \cdot \boldsymbol{x}: u(\boldsymbol{x}; \boldsymbol{z}, \boldsymbol{w}) \geq u\}$, and similarly, define $\boldsymbol{x}'$ to be the minimizer of $\min_{\boldsymbol{x}}\{\boldsymbol{p} \cdot \boldsymbol{x}: u(\boldsymbol{x}; t\boldsymbol{z}, \boldsymbol{w}) \geq tu\}$. Then,

$e(u, \boldsymbol{p}; \boldsymbol{z}, \boldsymbol{w}) = \min_{\boldsymbol{x}}\{\boldsymbol{p} \cdot \boldsymbol{x}: u(\boldsymbol{x}; \boldsymbol{z}, \boldsymbol{w}) \geq u\}$ [Definition]

$\Leftrightarrow e(u, \boldsymbol{p}; \boldsymbol{z}, \boldsymbol{w}) = \min_{\boldsymbol{x}}\{\boldsymbol{p} \cdot \boldsymbol{x}: tu(\boldsymbol{x}; \boldsymbol{z}, \boldsymbol{w}) \geq tu\}$

$\Leftrightarrow e(u, \boldsymbol{p}; \boldsymbol{z}, \boldsymbol{w}) = \min_{\boldsymbol{x}}\{\boldsymbol{p} \cdot \boldsymbol{x}: u(t\boldsymbol{x}; t\boldsymbol{z}, \boldsymbol{w}) \geq tu\}$ [Assumption: Joint Homogeneity of degree 1 in $u(\cdot)$]

$\Leftrightarrow e(u, \boldsymbol{p}; \boldsymbol{z}, \boldsymbol{w}) = \min_{\tilde{\boldsymbol{x}}}\{\boldsymbol{p} \cdot t^{-1}\tilde{\boldsymbol{x}}: u(\tilde{\boldsymbol{x}}; t\boldsymbol{z}, \boldsymbol{w}) \geq tu\}$ [Unit conversion: $\tilde{\boldsymbol{x}} = t\boldsymbol{x}$]

$\Leftrightarrow e(u, \boldsymbol{p}; \boldsymbol{z}, \boldsymbol{w}) = t^{-1}\min_{\tilde{\boldsymbol{x}}}\{\boldsymbol{p} \cdot \tilde{\boldsymbol{x}}: u(\tilde{\boldsymbol{x}}; t\boldsymbol{z}, \boldsymbol{w}) \geq tu\}$ [Min Operator Property]

$\Leftrightarrow \min_{\boldsymbol{x}}\{\boldsymbol{p} \cdot \boldsymbol{x}: u(\boldsymbol{x}; \boldsymbol{z}, \boldsymbol{w}) \geq u\} = t^{-1}\min_{\tilde{\boldsymbol{x}}}\{\boldsymbol{p} \cdot \tilde{\boldsymbol{x}}: u(\tilde{\boldsymbol{x}}; t\boldsymbol{z}, \boldsymbol{w}) \geq tu\}$ [Definition]

and therefore $\boldsymbol{x}^* = \boldsymbol{x}'$ since scaling an objective function by $t^{-1}$ does not affect the minimizer and the minimizers are unique by strict quasiconcavity. Alternatively, assuming a differentiable expenditure function, this corollary can be proven by differentiating the expenditure function and applying Shepard's Lemma. *Q.E.D.*